\documentclass{article} 
\usepackage{iclr2027_conference,times}

\usepackage{amsmath,amsfonts,bm}

\def\eqref#1{equation~\ref{#1}}

\def\1{\bm{1}}

\DeclareMathAlphabet{\mathsfit}{\encodingdefault}{\sfdefault}{m}{sl}
\SetMathAlphabet{\mathsfit}{bold}{\encodingdefault}{\sfdefault}{bx}{n}

\iclrfinalcopy
\usepackage{hyperref}
\usepackage{url}
\usepackage{booktabs}
\usepackage{multirow}
\usepackage{graphicx}
\usepackage{caption}
\usepackage{amsmath}

\usepackage{xcolor}
\usepackage{tikz}
\usetikzlibrary{arrows.meta,positioning}
\usepackage[shortcuts]{extdash}
\usepackage{comment}

\newcommand{\bciWorst}{15.6\%}
\newcommand{\bciBase}{90.8\%}
\newcommand{\obmiWorst}{9.9\%}
\newcommand{\obmiBase}{90.8\%}
\newcommand{\bciChance}{11.1\%}
\newcommand{\obmiChance}{1.85\%}
\newcommand{\bciCost}{7.0}
\newcommand{\obmiCost}{2.1}

\title{Riemannian Suppression of Cross-Session Identity Leakage in Motor-Imagery EEG}

\author{%
  Sheikh Samit Muhaimin, Aarthy Nagarajan \\
  Department of Computer Science and Engineering, University of Notre Dame \\
  \texttt{\{smuhaimi, anagaraj\}@nd.edu}
}

\renewcommand{\headrulewidth}{0pt}

\begin{document}

\maketitle

\begin{abstract}
Electroencephalography (EEG) recorded for motor-imagery (MI) brain--computer
interfaces (BCIs) carries subject-identifying structure that persists across sessions, allowing separately released sessions from the same person to be linked. Existing EEG identity protections use learned or additive perturbations and operate on representations other than the covariance geometry underlying MI pipelines. We ask whether cross-session identity leakage can be suppressed directly in this covariance representation without learning a privacy model. We introduce a Riemannian transform that equalizes two subject-conditioned geometric statistics:
the Riemannian mean covariance and the tangent-space scatter around it, with two
optional band-specific refinements. Ablations show that mean alignment and scatter whitening carry nearly all of the
suppression, while the optional stages contribute small, dataset-dependent
refinements. Across two public datasets and five attackers spanning four covariance-space
representations, the worst attacker falls from \bciBase\ to \bciWorst\ on BCI
Competition~IV-2a and from \obmiBase\ to \obmiWorst\ on OpenBMI, at a cost of \bciCost\ and \obmiCost\ points of MI decoding accuracy relative to the 8--30\,Hz input, and no aggregation rule exceeds chance by more than two points at any batch size. The
transform requires no learned privacy model, surrogate network, or
gradient-based training, and is fitted on one session and applied unchanged to a held-out later session.
\end{abstract}

\section{Introduction}
\label{sec:intro}

Motor-imagery (MI) brain--computer interfaces (BCIs) support communication and control for people with severe motor impairment \citep{wolpaw2002bci}, as well as neurological rehabilitation \citep{daly2008rehab}, robotic device control \citep{edelman2019noninvasive}, and awareness assessment in disorders of consciousness \citep{li2016multimodal}.
Advancing the field requires shared data, and the benchmarks supply it by
releasing every subject and session as recorded: BCI Competition~IV-2a
\citep{tangermann2012bci} publishes two sessions from each of nine subjects,
OpenBMI \citep{lee2019openbmi} from each of fifty-four. Cross-dataset transfer
learning depends on pooling them \citep{wu2022transfer}, and recent EEG
foundation models aggregate at far greater scale still
\citep{jiang2024labram}. This creates a risk distinct from the task the data was collected for:
EEG carries private information beyond the intended signal
\citep{ienca2018brain,xia2022privacy}. Resting-state EEG alone supports
subject identification at high accuracy \citep{choi2018individual}, deep
sequence models identify users from task recordings \citep{sun2019eeg}, and
side-channel work has shown that credentials and personal associations can be
inferred from BCI signals \citep{martinovic2012feasibility}. The specific risk we address is \emph{cross-session linkage}. Given two
separately released sessions, an adversary can determine which recordings
belong to the same individual, and thereby aggregate a person's data across
releases, sites, or time. \citet{meng2023identity} make the consequence
concrete: linked sessions make inference about health status or study
participation far more reliable than any single release permits, which
discourages participation in data collection.

Existing EEG privacy work addresses adjacent problems. Cryptographic and
secure-computation methods fit a model without any party seeing raw signals but never produce a shareable dataset. Perturbation-based methods do defend a release, but target a different representation and release model (Section~\ref{sec:related}). Our
setting differs on three counts. The transform requires no learned privacy model or gradient-based training;
its subject-specific transforms are estimated directly from covariance
statistics of a calibration session. Every released session is defended, so the adversary compares defended data
against defended data rather than a perturbed training set against a clean
test set: the protection is a property of each released recording, not of
what a classifier can be made to learn. And the representation is spatial covariance, which MI
decoding is built on and identity attackers read, yet no existing defense
targets.

The core difficulty is that identity and task occupy the same signal. MI decoding relies on event-related desynchronisation in the $\mu$ (8--13\,Hz) and $\beta$ (13--30\,Hz) rhythms, expressed through spatial covariance, while identity is encoded in the structure of that same covariance. A defense that removes one may therefore degrade the other. We hypothesize that persistent cross-session identity is carried by complementary subject-specific statistics, manifold location and tangent-space scatter, and ask whether jointly equalizing them can suppress identity while preserving task-relevant structure.

\paragraph{Contributions.} We identify subject-specific covariance mean and tangent-space scatter as complementary sources of cross-session identity leakage, and jointly equalize them with a Riemannian transform. Ablations
show they account for nearly all suppression, with band-specific stages providing only dataset-dependent refinements (Section~\ref{sec:ablation}). The pipeline substantially reduces five attackers spanning four
covariance-space representations on two public datasets, from over 90\% to
\bciWorst\ and \obmiWorst, at a measurable utility cost
(Section~\ref{sec:privacy}). At the session level, mean-tangent aggregation strengthens undefended linkage to 100\% and 90.7\%, but does not materially exceed chance after defense, no rule by more than two points (Section~\ref{sec:linkage}). Geometric alignment and prior perturbation-based protection address a different regime (Section~\ref{sec:baselines}).

\section{Related Work}
\label{sec:related}

Resting-state EEG supports subject identification at high accuracy
\citep{choi2018individual}, and identity persists in task-recorded EEG across
paradigms \citep{sun2019eeg}. \citet{meng2023identity} state the consequence
directly: identity is learned easily enough that separate sessions from one
user can be associated, making private inference more reliable than any
single release permits. Aggregation makes this worse, since LaBraM pre-trains on roughly 2{,}500 hours from around twenty datasets \citep{jiang2024labram} and the larger the pool, the more releases an adversary can cross-reference. Existing protections divide three ways. The closest prior work makes EEG ``identity-unlearnable'' by perturbing the training release, sample-wise or user-wise, so a user-classifier trained on it fails to generalise \citep{meng2023identity,huang2021unlearnable}; \citet{chen2025userwise} extend this to four perturbation types and \citet{meng2024multiprivacy} to gender and BCI-experience, while \citet{singh2023selective} train an autoencoder that outputs an anonymised signal and ID-RemovalNet \citep{wang2025idremovalnet} strips identity from learned features before decoding. All are evaluated against deep time-domain classifiers, and the perturbation methods pair a perturbed \emph{training} release with clean test data, a different attacker family and release model from ours. A parallel line makes the
data unreadable rather than unidentifying: secure multiparty computation fits
a model without any party observing raw signals
\citep{agarwal2019protecting}, and non-invertible neural encoding obscures
feature values while preserving class separability \citep{khanam2026dsnet};
both target confidentiality within a single session, and neither measures
re-identification or cross-session linkage. Third, Euclidean Alignment \citep{he2020ea} and related per-subject covariance normalisations, standard for cross-subject MI transfer, are sometimes assumed to confer privacy by construction. That assumption is only partly correct: EA suppresses re-identification substantially but
stops far short of chance (Section~\ref{sec:baselines}), so a practitioner
running it has protection and no way to know how much.

None of these targets the representation the attackers read. MI decoding is built on spatial covariance: Common Spatial Patterns and its filter-bank extension \citep{ang2008fbcsp} estimate spatial filters from class covariances, Riemannian tangent-space classification \citep{barachant2012riemannian} operates on the SPD manifold of trial covariances, and the alignment techniques behind cross-subject transfer \citep{zanini2018riemannian,he2020ea} are covariance transforms. A defense built elsewhere in the pipeline does not directly target these covariance-space attacks. Second-order and tangent-space alignment methods
\citep{sun2016coral,bleuze2022tsa} match distributions by whitening and
recolouring, but were developed for domain adaptation rather than
cross-session EEG identity suppression. Recentering and whitening are established components of domain and Riemannian alignment \citep{zanini2018riemannian, rodrigues2019rpa, bleuze2022tsa}, but these methods were developed for distribution alignment and transfer rather than for characterizing and suppressing cross-session identity leakage in EEG covariance geometry. 
\section{Threat Model}
\label{sec:threat}

A data owner holds recordings from $U$ subjects
across two or more sessions, and every released session is passed through the
defense. There is no clean copy behind it: the defended data \emph{is} the
data. Membership inference, and the poisoned-training-set setting of
\citet{meng2023identity} where only the training release is perturbed, are
different release models. The adversary sees only defended trials plus subject
labels for a reference session, trains a re-identification classifier from
scratch on it, then receives defended trials from a \emph{different} session
and links them to a subject. It never receives the fitted transform, and may
preprocess the release however it wishes. Our metric is re-identification accuracy on the held-out session, with
chance at $100/U$; utility is MI decoding accuracy on the same defended data.
Our privacy metric reports the worst case over the evaluated attackers, and therefore provides a lower bound on leakage. We measure it at two levels:
trial-level identification, the metric used throughout the covariance-attacker
literature, and session-level linkage, which aggregates the trials a subject
contributes into one decision. The second is what the threat model concerns, since evidence weak per trial may be decisive once pooled.

\section{Method}
\label{sec:method}

\begin{figure}[t]
\centering
\includegraphics[width=\textwidth]{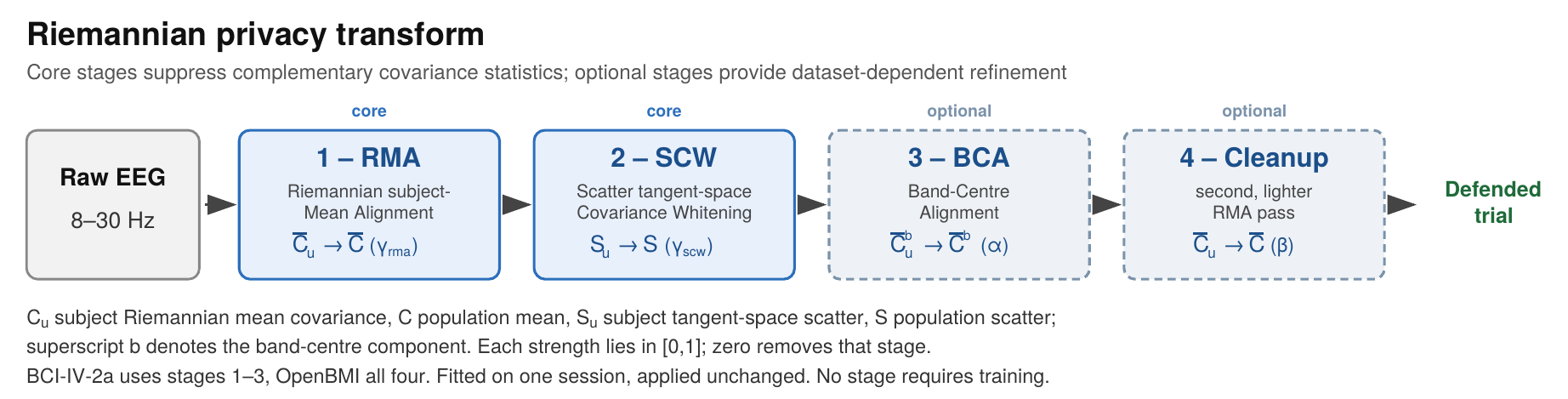}
\caption{The Riemannian privacy transform. Input is band-limited to
8--30\,Hz, then passed through up to four stages. The two core stages suppress complementary covariance statistics, RMA the subject mean covariance and SCW the tangent-space scatter about it, while BCA and the cleanup pass refine per dataset, realigning the band centre and correcting the drift it introduces. Each strength lies in
$[0,1]$ and zero removes its stage: BCI-IV-2a uses stages 1--3, OpenBMI all
four. The transform is fitted on one session and applied unchanged.}
\label{fig:pipeline}
\end{figure}

A raw trial is band-limited to 8--30\,Hz and passes through up to four alignment stages, emerging with the same shape, so every downstream decoder and attacker consumes it unchanged (Figure~\ref{fig:pipeline}). The two core stages equalize the subject mean covariance and tangent-space scatter. Two optional stages then refine residual band-specific structure and correct the resulting broadband drift. Each acts on the output of the previous
one. Let
$X_i \in \mathbb{R}^{c \times t}$ be a trial with $c$ channels and $t$
samples and $C_i = \frac{1}{t} X_i X_i^\top$ its spatial covariance,
regularised as $C_i + 10^{-6} I$. Every quantity the transform estimates, the
per-subject and population Riemannian means and the per-subject tangent-space
scatters, is estimated once from the fitting session. The
transform requires no learned privacy model, surrogate classifier, or
gradient-based optimisation. The released signal is band-limited to
8--30\,Hz by a Butterworth filter before the transform, as part of the method
rather than preprocessing. Band-limiting makes the attackers
\emph{stronger}, raising the undefended worst case from 85.1\% to 90.8\% on
BCI-IV-2a and from 89.2\% to 90.8\% on OpenBMI, so every defended number is
measured against the harder baseline and both are reported in
Table~\ref{tab:main}.

\paragraph{Stage 1: Riemannian subject-Mean Alignment (RMA,
$\gamma_{\mathrm{rma}}$).} Subjects differ first and most simply in where
their trials sit on the SPD manifold: each subject's covariances cluster
around a personal mean $\bar{C}_u$, and a classifier that learns those
locations has learned identity. Stage~1 moves every cluster onto the population mean $\bar{C}$. Write $L_A$
for the lower Cholesky factor of an SPD matrix $A$, so $A = L_A L_A^\top$.
Then $L_{\bar{C}_u}^{-1}$ carries $\bar{C}_u$ to the identity and
$L_{\bar{C}}$ carries the identity to $\bar{C}$; chaining the two gives a map
that lands the subject mean exactly on the population mean, and blending it
with the identity at strength $\gamma$ sets how far each subject moves:
\begin{equation}
W_u \;=\; L_{\bar{C}}\, L_{\bar{C}_u}^{-1},
\qquad
W_u(\gamma) \;=\; \gamma\, W_u \;+\; (1-\gamma)\, I .
\label{eq:rma-map}
\end{equation}
%

Applying $W_u(\gamma)$ to a covariance gives the target
$C_i^{\star} = W_u(\gamma)\, C_i\, W_u(\gamma)^{\!\top}$, but the transform
must return a \emph{signal}, not a covariance, so that downstream decoders
and attackers consume it unchanged. We therefore recolour the trial to have
that covariance,

\begin{equation}
\mathcal{R}(X_i, C^{\star}) \;=\; L_{C^{\star}}\, L_{C_i}^{-1} X_i ,
\label{eq:recolour}
\end{equation}
the unique lower-triangular map from $C_i$ to $C^{\star}$, whitening by $L_{C_i}^{-1}$ and recolouring by $L_{C^{\star}}$. At $\gamma = 0$ the stage is the identity; at $\gamma = 1$ each mean lands on $\bar{C}$.

\paragraph{Stage 2: Scatter tangent-space Covariance Whitening (SCW,
$\gamma_{\mathrm{scw}}$).} Aligning the means is not enough. Two subjects
whose covariances now share a centre can still be told apart by how their
trials \emph{spread} around it: the shape and orientation of that dispersion
is itself a subject signature. Stage~2 equalises the second moment. Dispersion is not well defined on a curved manifold, so we work in the
tangent space at $\bar{C}$, where the trials become ordinary vectors. Let
$\mathrm{vech}(\cdot)$ map the upper triangle of a symmetric $c \times c$
matrix to $\mathbb{R}^{d}$, $d = c(c{+}1)/2$, off-diagonals scaled by
$\sqrt{2}$ so the Euclidean inner product matches the Frobenius one. The
embedding of $C_i$ is
\begin{equation}
z_i \;=\; \mathrm{vech}\!\left(
  \log\!\left( \bar{C}^{-1/2} C_i \bar{C}^{-1/2} \right)\right)
  \in \mathbb{R}^{d},
\label{eq:tangent}
\end{equation}
with $\log$ the matrix logarithm. The population and per-subject scatters are
then just sample covariances of these vectors,
\begin{equation}
S \;=\; \mathrm{Cov}\big(\{z_i\}_{i}\big) + \lambda I ,
\qquad
S_u \;=\; \mathrm{Cov}\big(\{z_i : u_i = u\}\big) + \lambda I ,
\label{eq:scatter}
\end{equation}
with $\lambda = \lambda_{\mathrm{SCW}} = 10^{-4}$.
Estimating $S_u$ requires shrinkage toward the scaled identity whenever $n_u \le d$, always the case on OpenBMI; Appendix~\ref{app:scw} gives the form and constant.
The whitening map takes each subject's scatter to the population scatter,
\begin{equation}
A_u \;=\; S^{1/2} S_u^{-1/2},
\qquad
A_u(\gamma) \;=\; \gamma\, A_u + (1-\gamma)\, I ,
\label{eq:scw-map}
\end{equation}
so at $\gamma = 1$ every subject's trials have the same second moment about
the shared mean. Each embedding is mapped $z_i \mapsto A_u(\gamma) z_i$ and
returned to the manifold by inverting \eqref{eq:tangent},
\begin{equation}
C_i^{\star} \;=\; \bar{C}^{1/2}
  \exp\!\big(\mathrm{vech}^{-1}(A_u(\gamma) z_i)\big) \bar{C}^{1/2},
\label{eq:scw-back}
\end{equation}
and recoloured by \eqref{eq:recolour}; tangent eigenvalues are clipped before the exponential (Appendix~\ref{app:scw}). This whitening--recolouring step is analogous in form to second-order
covariance alignment, but is applied here to subject-conditioned
tangent-space scatter as a privacy transform. Both $\bar{C}$ and the maps
$A_u$ are estimated once on the fitting session and reused unchanged, so the
tangent point does not move between sessions. This is what makes the
transform a fixed map that can be applied to future releases rather than a
per-session recomputation.


\paragraph{Stage 3: Band-Centre Alignment (BCA, $\alpha$).} Stages 1--2 treat
the 8--30\,Hz covariance as a single object, so they equalise it on average
across the band. Identity that lives disproportionately where the motor
rhythms are strongest is only partly reached by that average. Stage~3 applies
the same alignment a second time, restricted to the band centre. Let $X_{b,i}$ be the output of the same fifth-order Butterworth band-pass
applied a second time to the already band-limited trial. The second pass
attenuates the band edges and leaves a narrower core carrying roughly 88\% of
the in-band energy. Stage~1 is refitted on $\{X_{b,i}\}$ alone, giving maps $W^b_u(\alpha)$ by \eqref{eq:rma-map}, and the trial becomes
\begin{equation}
\tilde{X}_i \;=\; (X_i - X_{b,i}) \;+\;
  \mathcal{R}\big(X_{b,i},\; W^b_u(\alpha)\, C_{b,i}\, W^b_u(\alpha)^{\!\top}\big).
\label{eq:bca}
\end{equation}
The residual $X_i - X_{b,i}$ is added back untouched, so the stage
concentrates alignment on the band centre and leaves the edges as they were.
Section~\ref{sec:ablation} shows it is marginal and dataset-dependent.

\paragraph{Stage 4: cleanup RMA ($\beta$).} Because Stage~3 modifies only part of the spectrum, the broadband covariance Stages 1--2 equalised is no longer exactly equalised. Stage~4 corrects that drift: it
refits \eqref{eq:rma-map} on the Stage-3 output at strength $\beta$ and
applies \eqref{eq:recolour} again.

\paragraph{Configuration.} RMA and SCW form the core transform, while BCA and cleanup are optional refinements whose strengths may be set to zero. BCI-IV-2a uses BCA only, while OpenBMI uses both. Section~\ref{sec:ablation} shows that RMA and SCW account for nearly all suppression, with the refinements shifting the worst attacker by less than one point.
The privacy stages use fifth-order Butterworth band-pass filters in
second-order-section form, applied with zero phase, while the MI decoding
baselines use the reference FBCSP toolbox Chebyshev~II filter bank;
Appendix~\ref{app:filter} shows the two are equivalent across the 8--30\,Hz
band and records which design each component uses.

\section{Experimental Setup}
\label{sec:setup}


\paragraph{Datasets and protocol.} \emph{BCI Competition IV-2a}: 9 subjects, 22 channels, 4-class MI (left hand, right hand, feet, tongue), 2 sessions, 288 trials each, 250\,Hz. Chance re-identification \bciChance; chance MI 25\%.
\emph{OpenBMI}: 54 subjects, 20 channels (motor montage), 2 sessions,
resampled to 250\,Hz. Chance re-identification \obmiChance. The transform is fit on Session~1, the attacker trained on defended Session~1
and tested on defended Session~2. All primary results are
cross-session; the single-session baseline of Appendix~\ref{app:bakjeong} is
separate. We evaluate
five covariance-space attackers across four representations: SVM (broadband covariance), XDAWN+SVM (evoked-augmented covariance), SPDNet-ID (deep SPD-manifold network),
CSP-SVM and CSP-LDA (narrow-band spatial patterns).

\paragraph{Attacker configuration.} Attacker configurations are fixed
independently of Session~2 except for XDAWN, reporting an oracle
maximum over a predefined set of filter counts,
$\mathrm{nfilter} \in \{1,2,4,8\}$. This oracle selection provides a
conservative upper bound on XDAWN attacker performance by giving the attacker the best preprocessing configuration on the held-out session;
the selected count is 8 undefended and 1 defended on BCI-IV-2a.
Tangent-space SVM: OAS covariance, Riemannian tangent space, RBF SVC at
scikit-learn defaults. XDAWN+SVM: OAS-estimated \texttt{XdawnCovariances}
with the evoked filters applied. CSP-SVM and CSP-LDA:
one-versus-rest CSP on trace-normalised covariances, then an RBF SVC and an
LDA respectively on the same features. SPDNet-ID: BiMap dimensions
$c \to c/2 \to c/4$, 400 epochs, Adam at $5\times10^{-4}$, batch 64, early
stopping at patience 40 on a 10\% stratified split, seeded to 42. SPDNet-ID tracks label-space size rather than signal, falling from 87.0\% on 9 OpenBMI subjects to 66.1\% on all 54 (Appendix~\ref{app:spdnet}).

\paragraph{Utility decoders.} FBCSP \citep{ang2008fbcsp}, EEGNet
\citep{lawhern2018eegnet}, DeepConvNet \citep{schirrmeister2017deep}, and
FBCNet \citep{mane2021fbcnet}, trained and evaluated on the same defended
data. One utility decoder was fixed per dataset before operating-point
selection and held fixed throughout: FBCNet on BCI-IV-2a, EEGNet on OpenBMI.
The admissibility floor and every reported cost therefore use the same
architecture, measured against that decoder's own undefended accuracy. All
four decoders are reported in Table~\ref{tab:main}.

\paragraph{Operating-point selection.} Session~1 is split per subject by
trial order into an inner enrolment and query set. The transform is fitted on enrolment data, and privacy and utility are evaluated on the
query split, leaving Session~2 untouched. Let $u_0$ be the decoder's undefended query accuracy. A grid cell is admissible if its
decoder accuracy is at least $u_0 - s$. Among admissible cells, we select
the one with the lowest worst-case attacker accuracy. The grid is
$\gamma_{\mathrm{rma}} \in \{0.8, 0.9, 1.0\}$,
$\gamma_{\mathrm{scw}} \in \{0.35, 0.5, 0.6, 0.7, 0.8\}$,
$\alpha \in \{0.3, 0.6\}$ and $\beta \in \{0, 0.2\}$. The
selected setting is refit on full Session~1 and evaluated once on
Session~2. The utility slack $s$ is the one free choice in the procedure. We use $s = 0.5$ for BCI-IV-2a and $s = 2.5$ for OpenBMI. Appendix~\ref{app:sensitivity} gives the full sensitivity analysis. 

\section{Covariance-Space Privacy Results}
\label{sec:privacy}

%
%

\begin{table}[t]
\centering
\caption{Cross-session re-identification and MI decoding (\%) at the reported
operating points. XD = XDAWN+SVM, SPD = SPDNet-ID, CSP-S/L = CSP-SVM/LDA.
Worst is over the five attackers. The fixed utility decoder, FBCNet on
BCI-IV-2a and EEGNet on OpenBMI, is shown in bold, so every cost is
within-decoder. $\Delta$ is defended minus band-limited. Chance is
\bciChance\ and \obmiChance\ for re-identification, 25\% and 50\% for
decoding.}
\label{tab:main}
\scriptsize
\setlength{\tabcolsep}{4.2pt}
\begin{tabular}{lcccccc@{\hskip 10pt}cccc}
\toprule
& \multicolumn{6}{c}{Re-identification $\downarrow$}
& \multicolumn{4}{c}{MI decoding $\uparrow$} \\
\cmidrule(lr){2-7}\cmidrule(lr){8-11}
 & SVM & XD & SPD & CSP-S & CSP-L & Worst
 & FBCSP & EEGNet & DCN & FBCNet \\
\midrule
\multicolumn{11}{l}{\emph{BCI-IV-2a}, $(\gamma_{\mathrm{rma}},\gamma_{\mathrm{scw}},\alpha,\beta) = (0.9,0.35,0.6,0.0)$} \\
\quad No defense, raw       & 75.9 & 85.1 & 67.4 & 72.2 & 77.9 & 85.1
                            & 61.9 & 67.3 & 70.1 & 73.6 \\
\quad No defense, 8--30\,Hz & 90.8 & 87.6 & 89.9 & 73.2 & 77.3 & 90.8
                            & 58.0 & 61.1 & 57.3 & 71.9 \\
\quad Defended, S1$\to$S2   & 11.1 & 15.5 & 15.1 & 15.6 & \phantom{0}3.5 & \textbf{15.6}
                            & 49.0 & 57.8 & 52.3 & \textbf{64.9} \\
\quad Defended, S2$\to$S1   & 11.1 & 12.9 & 12.1 & \phantom{0}5.8 & \phantom{0}0.9 & \textbf{12.9}
                            & 48.8 & 56.0 & 52.6 & \textbf{66.5} \\
\quad $\Delta$ (S1$\to$S2)  & $-79.7$ & $-72.1$ & $-74.8$ & $-57.6$ & $-73.8$ & $-75.2$
                            & $-9.0$ & $-3.3$ & $-5.0$ & $-7.0$ \\
\midrule
\multicolumn{11}{l}{\emph{OpenBMI}, $(\gamma_{\mathrm{rma}},\gamma_{\mathrm{scw}},\alpha,\beta) = (0.9,0.60,0.6,0.2)$} \\
\quad No defense, raw       & 89.2 & 79.7 & 40.4 & 83.3 & 86.0 & 89.2
                            & 61.8 & 64.5 & 63.5 & 67.5 \\
\quad No defense, 8--30\,Hz & 90.8 & 88.6 & 66.1 & 83.5 & 85.9 & 90.8
                            & 65.6 & 65.1 & 62.6 & 64.9 \\
\quad Defended, S1$\to$S2   & \phantom{0}2.4 & \phantom{0}9.9 & \phantom{0}1.5 & \phantom{0}2.8 & \phantom{0}3.8 & \textbf{\phantom{0}9.9}
                            & 60.1 & \textbf{63.0} & 60.5 & 58.0 \\
\quad Defended, S2$\to$S1   & \phantom{0}2.2 & \phantom{0}9.1 & \phantom{0}2.2 & \phantom{0}2.2 & \phantom{0}3.6 & \textbf{\phantom{0}9.1}
                            & 61.1 & \textbf{62.3} & 59.0 & 54.2 \\
\quad $\Delta$ (S1$\to$S2)  & $-88.4$ & $-78.7$ & $-64.6$ & $-80.7$ & $-82.1$ & $-80.9$
                            & $-5.5$ & $-2.1$ & $-2.1$ & $-6.9$ \\
\bottomrule
\end{tabular}
\end{table}

Table~\ref{tab:main} reports the headline cross-session result. The worst
attacker falls from \bciBase\ to \bciWorst\ on BCI-IV-2a and from \obmiBase\
to \obmiWorst\ on OpenBMI, at a cost of \bciCost\ and \obmiCost\ points of
decoding accuracy. The cost is not uniform across decoders. On BCI-IV-2a the
filter-bank decoders lose most, FBCSP $-9.0$ and FBCNet $-7.0$ points, while
the single-band deep decoders lose five or fewer. On OpenBMI the same pattern
holds more weakly, FBCSP $-5.5$ and FBCNet $-6.9$ against $-2.1$ for EEGNet
and DeepConvNet. FBCNet is strongest on BCI-IV-2a but weakest on OpenBMI,
where the two-class task and twenty-channel montage give a filter bank less
to exploit. Appendix~\ref{app:erds} shows the corresponding
physiological cost: $\beta$-band desynchronisation survives with its
topography correlating at $r = 0.79$, but the C3--C4 gradient compresses from $-5.04$ to $-1.16$ points.

\subsection{Privacy--utility trade-off}

\begin{figure}[t]
\centering
\includegraphics[width=0.82\textwidth]{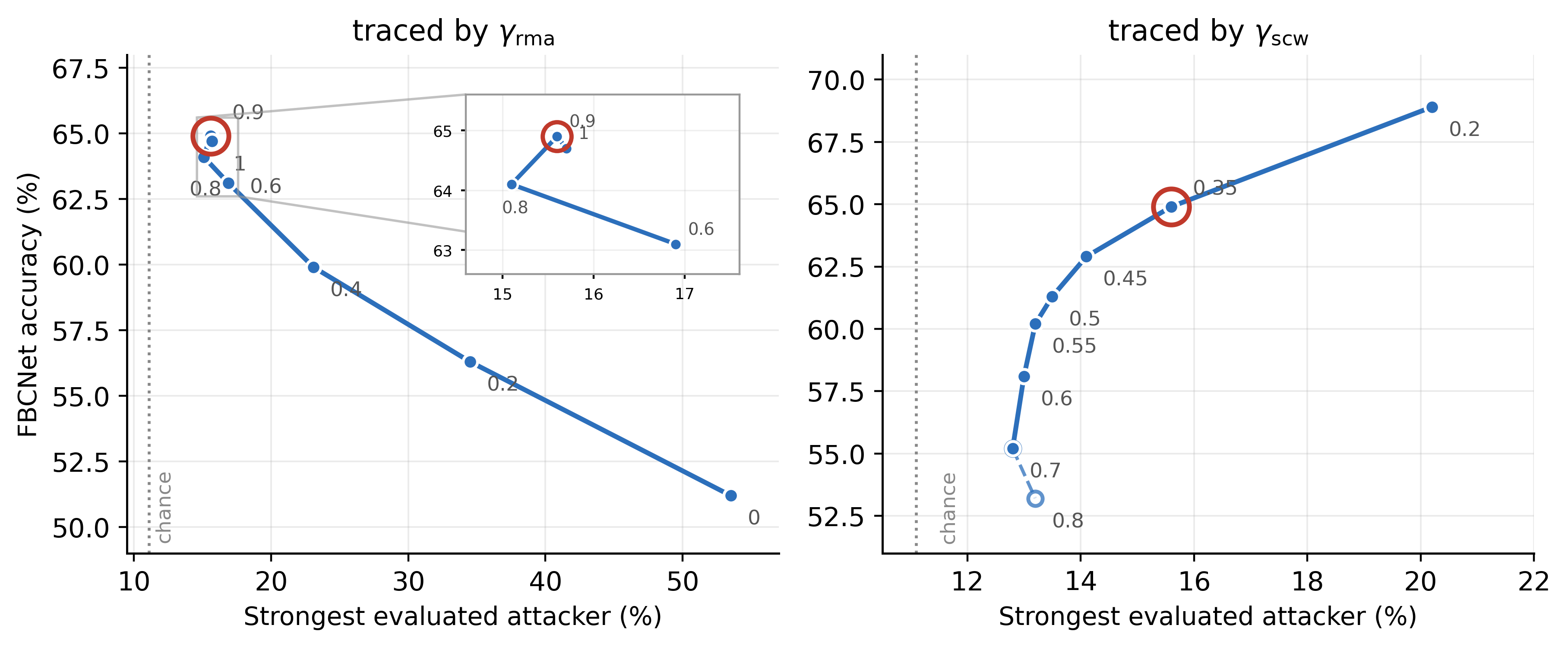}
\caption{Privacy--utility curves on BCI-IV-2a, varying one strength with the
other three at $(0.9,0.35,0.6,0.0)$; the circle marks the selected point.
Left: $\gamma_{\mathrm{rma}}$ moves both axes together, so under-aligning
costs privacy and accuracy at once; the inset expands the flat tail. Right:
$\gamma_{\mathrm{scw}}$ trades roughly a point of accuracy per point of
privacy, and the open marker at $0.80$ leaves the frontier. CSP-SVM binds
throughout; chance is \bciChance\ (dotted).}
\label{fig:opcurve}
\end{figure}

Figure~\ref{fig:opcurve} traces the trade-off on BCI-IV-2a; the two strengths
behave differently. $\gamma_{\mathrm{scw}}$ traces a
frontier: the worst attacker falls 20.2\%, 15.6\%, 14.1\%, 13.5\%, 13.2\%,
13.0\% and 12.8\% as it runs 0.20 to 0.70, while FBCNet falls 68.9\% to
55.2\%; roughly a point of decoding accuracy per point of privacy. The
selected setting sits at $\gamma_{\mathrm{scw}} = 0.35$, where the curve is
still steep and the decoding cost is \bciCost\ points.
$\gamma_{\mathrm{rma}}$ does not trade at all over most of its range: raising
it from 0 to 0.9 improves the worst attacker from 53.5\% to \bciWorst\
\emph{and} FBCNet from 51.2\% to 64.9\%. Under-aligning is worse on both axes, and only past 0.9 does the curve flatten, with 1.0 giving 15.7\% and 64.7\%.

\paragraph{What selection costs.} A defender cannot see the evaluation
session, but if they could, the best would be $(0.8,0.35,0.6,0.2)$ at
15.0\% worst-case and 64.3\% FBCNet. Selecting on Session~1 gives
$(0.9,0.35,0.6,0.0)$ at \bciWorst\ and 64.9\%: 0.6 points worse on privacy,
0.6 better on accuracy. Neither point dominates the other, so the protocol costs nothing but a position on the same frontier. 

\subsection{Session-level linkage}
\label{sec:linkage}

An adversary holding a released session holds every trial in it, and evidence
weak per trial may become decisive once pooled. We therefore build subject
templates from defended Session~1 and aggregate $N$ defended Session~2 trials
into one decision under four rules (Appendix~\ref{app:linkage}). Aggregation
does help an undefended adversary: under the mean-tangent rule, accuracy rises from 88.3\% at one trial to 100\% at fifty on BCI-IV-2a, and from 82.5\% to 90.7\% on
OpenBMI, so trial-level accuracy understates undefended risk. Under defense no aggregation rule exceeds chance by more than two percentage
points at any batch size. The mean-tangent rule in fact \emph{declines} as more trials are added, because the residual variation the transform leaves is centred on the population mean rather than a subject-specific mean. The geometry says the same
without a classifier: same-subject cross-session distance relative to
cross-subject distance rises from 0.47 to 1.05 and from 0.37 to 1.00, so
after the transform two sessions from one person are no closer than two from
different people. Aggregation therefore reaches chance where trial-level identification does not, which is where the linkage threat is defined.

\subsection{Identity and task, measured the same way}
\label{sec:separation}

\begin{figure}[t]
\centering
\includegraphics[width=0.85\textwidth]{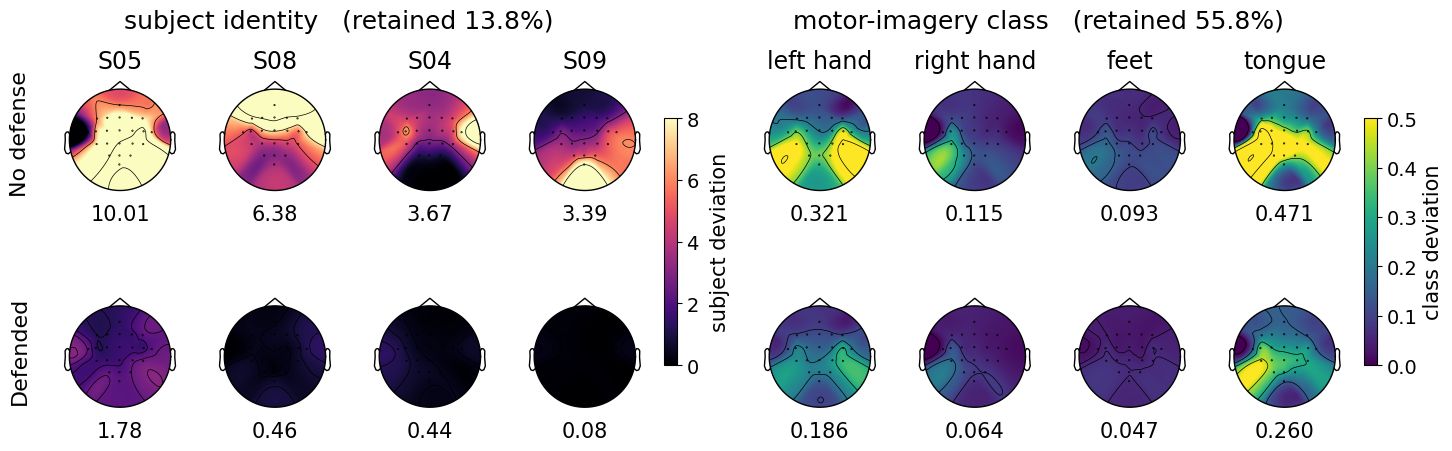}
\caption{Subject identity and motor-imagery class, measured the same way,
Session~2 on BCI-IV-2a. Separate colour scales: class differences in band
power are far smaller than differences between people, 0.25 against 3.24
undefended. Class deviation is uneven, tongue and left hand being more
visible in per-channel power than right hand or feet, which is why decoders
use spatial filters. The four subjects shown are the most identifiable of the
nine.}
\label{fig:sep}
\end{figure}

Identity and task can be compared with one instrument: the squared deviation
of a group mean log band power from its reference, in units of the relevant
within-group trial variance. Under the transform, subject deviation retains
13.8\% on BCI-IV-2a against 55.8\% for class deviation, and 20.9\% against
44.3\% on OpenBMI. Under this matched band-power measure, identity-related structure is therefore attenuated more strongly than task-related structure (Figure~\ref{fig:sep}; Appendix~\ref{app:sep}).

\subsection{Where the transform stops}
\label{sec:boundary}

All attackers above read spatial covariance, the object the transform
manipulates, so we also evaluate two that do not: per-channel band power over
8--30\,Hz with an RBF SVM, which discards cross-channel structure, and
EEGNet trained for subject identification. Both fall on both datasets. On
BCI-IV-2a bandpower drops from 84.3\% to 72.7\% and EEGNet-ID from 94.4\% to
83.3\%, on OpenBMI from 69.0\% to 33.6\% and 79.1\% to 38.1\%. These attacks
remain far above chance, so we scope the claim to covariance-space
protection. The deep time-domain case is complementary to \citet{meng2023identity}, as examined in Section~\ref{sec:baselines}.

\section{Mechanistic Ablation}
\label{sec:ablation}

We measure ten configurations at each reported operating point, removing a stage or tying the broadband strengths (Table~\ref{tab:ablation}). No stage suffices alone, and the three single-stage rows leave the worst attacker between 29.5\% and 84.7\%. RMA
is load-bearing on both axes, since removing it raises the worst attacker to
53.5\% and 61.4\% while \emph{also} costing 13.7 and 6.5 points of accuracy;
SCW is the clean trade, recovering 5.9 and 1.8 points of accuracy at the cost
of 22.5 and 53.1 points of privacy.

\begin{table}[t]
\centering
\caption{Stage ablation, cross-session S1$\to$S2, at each reported operating
point. Privacy is the worst of five attackers; utility is FBCNet on BCI-IV-2a
and EEGNet on OpenBMI. Tied fixes $\gamma_{\mathrm{rma}} =
\gamma_{\mathrm{scw}}$. Chance \bciChance\ and \obmiChance.}
\label{tab:ablation}
\small
\setlength{\tabcolsep}{5pt}
\begin{tabular}{lcccc}
\toprule
& \multicolumn{2}{c}{BCI-IV-2a} & \multicolumn{2}{c}{OpenBMI} \\
\cmidrule(lr){2-3}\cmidrule(lr){4-5}
Configuration & Worst $\downarrow$ & FBCNet $\uparrow$
              & Worst $\downarrow$ & EEGNet $\uparrow$ \\
\midrule
Selected configuration & \textbf{15.6} & \textbf{64.9} & \textbf{\phantom{0}9.9} & \textbf{63.1} \\
\midrule
No RMA ($\gamma_{\mathrm{rma}}{=}0$) & 53.5 & 51.2 & 61.4 & 56.6 \\
No SCW ($\gamma_{\mathrm{scw}}{=}0$) & 38.1 & 70.8 & 63.0 & 64.9 \\
No BCA ($\alpha{=}0$)                & 15.3 & 65.3 & 10.5 & 63.3 \\
No cleanup ($\beta{=}0$)             & 15.6 & 64.9 & 10.1 & 63.1 \\
\midrule
RMA only & 29.5 & 71.0 & 59.0 & 65.0 \\
SCW only & 84.1 & 53.3 & 84.7 & 58.8 \\
BCA only & 45.3 & 71.2 & 65.8 & 64.8 \\
\midrule
Tied, low  (0.35 / 0.60) & 25.3 & 58.4 & 14.6 & 62.0 \\
Tied, high (0.90 / 0.90) & 14.0 & 50.0 & \phantom{0}6.7 & 60.8 \\
\bottomrule
\end{tabular}
\end{table}

The mechanism is measurable in the tangent space the attackers read. On Session~2 the distance between subject means barely moves, 13.04 to 12.41 on BCI-IV-2a and 16.15 to 17.12 on OpenBMI, while the spread of a subject's trials about their mean roughly doubles, 8.03 to 14.69 and 10.11 to 26.60, so the within-to-between ratio rises to 1.19 and 1.56. The transform inflates each subject's dispersion rather than pulling subjects together, and Section~\ref{sec:separation} shows this per channel.

BCA is marginal on both datasets: removing it moves the worst attacker by 0.3 points on BCI-IV-2a and 0.6 on OpenBMI, and $\beta$ behaves the same way, selected at or beside the minimum on both. Only full strength is harmful for either, CSP-LDA rising to 13.7\% at $\alpha=1$ and 18.7\% at $\beta=1$. Tying the two
broadband strengths is worse on BCI-IV-2a, but on OpenBMI the tied setting at
$0.9$ reaches 6.7\%, below \obmiWorst, at 2.3 points of EEGNet accuracy;
adopting it on a cross-session measurement would be the test-set selection
the protocol exists to avoid.
Appendix~\ref{app:sensitivity} sweeps all four strengths on both datasets, and a $4^4$ sweep of 256 configurations bounds what blind selection gives up.
\section{Baseline Comparisons}
\label{sec:baselines}

\paragraph{Geometric alignment is a weak de-identifier.} Under an identical
protocol, Euclidean Alignment reduces the worst attacker from \bciBase\ to
69.0\% on BCI-IV-2a and from \obmiBase\ to 48.4\% on OpenBMI while
\emph{raising} decoding accuracy (Table~\ref{tab:app-ea}). That combination is the signature of a transfer normalisation: it removes between-subject variation, where identity partly lives. A recentring-and-scaling baseline after Riemannian Procrustes Analysis (RPA) \citep{rodrigues2019rpa} goes further, recentring each subject to the identity and rescaling their dispersion by a scalar where SCW whitens the tangent space. It is not RPA as published: we omit the rotation step, which \citet{rodrigues2019rpa} report barely improves MI classification, and generalise from the pairwise form to a cohort target. We write RPA-RS for this variant. It leaves the worst attacker at 46.3\% and 54.5\%, bound by XDAWN or tangent-space SVM rather than CSP attackers it suppresses (Table~\ref{tab:app-ea}). On BCI-IV-2a it buys that at a smaller decoding cost than ours, so the two sit on one frontier; on OpenBMI it does not, reading 63.6\% EEGNet against our 63.1\% (Table~\ref{tab:app-ea}) while leaving five times the leakage. Neither is adequate: six and twenty-six times chance for EA, four and twenty-nine for RPA-RS. A practitioner running either for transfer has protection they did not ask for and cannot quantify.

\paragraph{A published attack, on its own terms.} The five attackers of
Section~\ref{sec:setup} are constructed for this paper.
Appendix~\ref{app:bakjeong} reproduces the single-session identification
method of \citet{bak2023identification} on OpenBMI at 99.78\% against their
reported 98.97\%, and drives it to HTER 0.494 on the 20-channel montage, approximately chance-level, and 0.430 on a 36-channel reconstruction.

\paragraph{Perturbation-based protection covers a different representation.}
\citet{meng2023identity} add a learned user-wise perturbation to each
subject's trials so that a user-classifier trained on the perturbed release
fails to generalise. We reproduce it and run each method against the other's attackers (Appendix~\ref{app:meng}). Their perturbation reduces deep time-domain identification sharply, EEGNet from 68.7\% to 8.5\% and ShallowConvNet to 19.9\%, and leaves the covariance attackers untouched, the worst of our five staying at \bciBase\ and none moving by more than 0.3 points. Ours is the mirror image, reducing
all five covariance attackers to \bciWorst\ while leaving EEGNet
identification at 83.3\% (Section~\ref{sec:boundary}). The two are therefore
complementary rather than competing, and built for different release models:
ours defends every released session, theirs perturbs a training release
evaluated against clean data. Whether a single release can be protected in
both representations at once is open.

\paragraph{Differential privacy.} To our knowledge, existing DP work in BCI has focused on privacy during model training rather than release-time protection against cross-session EEG re-identification \citep{xia2022privacy,fu2025dpmld}. Under our trial-covariance release formulation, standard Laplace and Gaussian mechanisms \citep{dwork2014algorithmic} suppress covariance-space attackers but reduce MI decoding to near-chance performance: worst-case re-identification falls to 21.7--34.2\% in central mode and 28.4--29.6\% in local mode, while decoder accuracy remains only 29\%--36\% against 25\% chance, compared with 64.9\% under our transform. Under the stated adjacency, the calibrated noise dominates the covariance signal ($\Delta_1 = 44$ versus entries of order $1/22$),
and a ten-fold increase in $\varepsilon$ changes neither axis materially. Thus, under this direct covariance-release formulation, the calibrated DP mechanisms do not retain useful MI accuracy. Appendix~\ref{app:dp} gives the formulation and full results. Central mode was evaluated in full on BCI-IV-2a and, given the resulting utility collapse, was not repeated on OpenBMI.

\section{Statistical Significance}
\label{sec:stats}
Appendix~\ref{app:stats} reports subject-level
paired differences with bootstrap 95\% CIs, Cohen's $d_z$, and Wilcoxon
signed-rank tests against the band-limited baseline. All five covariance-space attackers decrease significantly on both datasets, by 57.6--88.4 points, with large effects. Utility losses are much smaller and less consistent: on BCI-IV-2a, FBCSP and FBCNet decrease significantly ($p = 0.039$ and $0.004$), whereas EEGNet and DeepConvNet do not ($p = 0.16$ and $0.20$).
\section{Conclusion}
\label{sec:conclusion}

Subject identity in motor-imagery EEG survives across recording sessions in
the spatial covariance that classical BCI pipelines consume. Our results show that this persistent identity is strongly expressed through two geometric properties of this representation: the subject-specific covariance mean and the tangent-space scatter around it. A Riemannian transform that jointly equalizes these statistics reduces five attackers spanning four covariance-space representations to \bciWorst\ and \obmiWorst\ on two public datasets, at costs of
\bciCost\ and \obmiCost\ points for the fixed utility decoder on each
dataset. Ablations show that mean alignment and scatter whitening are the load-bearing components, with the remaining stages providing only small dataset-dependent refinements. The transform is fitted on one session and applied unchanged to a held-out later session, and suppresses covariance-space identity substantially more strongly than standard alignment baselines.

The result is deliberately scoped. The transform does not provide universal EEG anonymization: attackers operating directly on time-domain signals or per-channel power retain substantial identity information. Rather, it establishes that cross-session covariance-space identity can be strongly suppressed with a fixed geometric map while preserving considerably more task information. Future work should determine whether complementary mechanisms can jointly suppress identity across covariance and time-domain representations, and evaluate linkage between independently held-out future sessions and under stronger adversarial knowledge of the fitted transform.

\section*{Ethics Statement}

This work studies re-identification of individuals from publicly released EEG
and proposes a defense against it. Both datasets are public and were
collected under their own institutional approvals
\citep{tangermann2012bci,lee2019openbmi}; we introduce no new recordings and
no new subjects. The attackers we construct are evaluated only on these
public releases and are reported so that the defense can be measured against
them, not as tools for identifying individuals. Section~\ref{sec:boundary}
states the limits of the protection explicitly, since a defense that is
believed to be stronger than it is would be worse than none: attackers
reading the released signal outside the covariance representation retain
substantial identification ability, and we do not claim protection there.

\section*{AI Use Statement}

In this work, we used generative AI tools for correcting the grammatical errors and punctuation of this paper, and for consistency checking against our own result files. All research questions, experimental design, analysis and conclusions are the authors' own, and the authors take full responsibility for the content.

\section*{Reproducibility Statement}

The transform is specified in full by Equations~(1)--(7), the shrinkage of
Appendix~\ref{app:scw}, and the constants in
Section~\ref{sec:setup}. The operating-point selection procedure, including
the grid and the per-dataset utility slack, is stated in
Section~\ref{sec:setup}; the selected settings are
$(0.9,0.35,0.6,0.0)$ on BCI-IV-2a and $(0.9,0.60,0.6,0.2)$ on OpenBMI.
Attacker configurations, including the SPDNet-ID architecture and seed, are
given in the same section. Both datasets are public
\citep{tangermann2012bci,lee2019openbmi}. Every appendix table states the
configuration it was measured at, and the two baseline reproductions
(Appendices~\ref{app:meng} and \ref{app:bakjeong}) report our figures beside
the published ones. Source code, together with the result file behind each reported table, will be released publicly. The Session-1 selection grids are
included as JSON, one row per cell with the inner worst-case attacker and
decoder accuracy, so the reported operating points can be recovered from the
stated rule without rerunning the transform. The covariance attackers are
deterministic given the data and reproduce exactly; the MI decoders are not,
and repeated runs of one configuration differ by a few tenths of a point, occasionally more for the deep decoders (Appendix~\ref{app:opnote}).

\bibliography{references}
\bibliographystyle{iclr2027_conference}

\appendix


\section{Note on Operating Points}
\label{app:opnote}
Every table in this appendix states the configuration it was measured at.
Sweeps and ablations are mechanism and sensitivity analyses; the operating
points reported in the main text were selected on a Session-1-only split, and
no reported point was chosen on the basis of a test-session result. The MI decoders are stochastic, so a decoder cell measured in two experiments differs between runs: Table~\ref{tab:main} and Appendix~\ref{app:stats} are separate runs, and their decoder cells differ by a few tenths of a point for the filter-bank decoders and by up to about 1.5 points for DeepConvNet, the most variable. Each table reports its own experiment's run.

\section{SPDNet-ID Cohort Size}
\label{app:spdnet}

SPDNet-ID reads 89.9\% undefended on BCI-IV-2a but 66.1\% on OpenBMI, which
looks like less leakage on the larger cohort. Sweeping OpenBMI at 9, 18,
27 and 54 subjects gives 87.0\%, 82.4\%, 74.0\% and 66.1\%, so the gap is
label-space size rather than a difference in what the signal carries.

\section{Numerical Details of Stage 2}
\label{app:scw}

The tangent embedding has $d = c(c{+}1)/2$ dimensions but a subject
contributes only $n_u$ trials, and on OpenBMI $d = 210$ against $n_u = 200$:
$S_u$ is singular by construction, and inverting it amplifies eigenvalues
that are numerical noise. Whenever $n_u \le d$ we therefore shrink toward the
scaled identity before inversion,
\begin{equation}
S_u \;\leftarrow\; (1-r)\, S_u \;+\; r\,\frac{\mathrm{tr}(S_u)}{d}\, I ,
\qquad r = 0.1 ,
\label{eq:shrink}
\end{equation}
and additionally clip eigenvalues of $S_u$ below
$10^{-8}\lambda_{\max}(S_u)$. Without this the inverse square root reaches
$10^{6}$ and the stage destroys the signal rather than whitening it.

Tangent eigenvalues are clipped to $[-40, 40]$ before the exponential of \eqref{eq:scw-back}: on OpenBMI $\|A_u\|$ approaches that magnitude and an unclipped
$\exp$ overflows to $10^{20}$ from a raw signal scale of $10^{3}$.

\paragraph{Cost.} Fitting and applying the transform to both sessions of
BCI-IV-2a, 5{,}184 trials at 22 channels and 1{,}000 samples, takes 18\,s on
one CPU core, 3.4\,ms per trial. Subsampling the montage to 8, 12 and 16
channels gives 8.2, 10.1 and 13.3\,s, so the cost grows close to linearly in
channel count over this range: it is dominated by the per-trial recolouring
rather than by the per-subject matrix decompositions.

\paragraph{Linear versus geodesic blending.} The strength blends in Equations~\ref{eq:rma-map} and \ref{eq:scw-map} are linear in the ambient
matrix space rather than geodesic on the manifold. The two agree at the
endpoints and differ in between; we use the linear form because it is what
the reported results were produced with, and note the distinction because the
choice is not neutral at intermediate strength.
\section{Alignment Baselines}
\label{app:ea}

Euclidean Alignment (EA) whitens each subject's trials by their mean covariance, without reference
to any other subject. It is standard practice in cross-subject MI transfer
and is not proposed as a privacy mechanism; we evaluate it because
practitioners already run it and may assume it protects identity. RPA-RS recentres each subject to the identity and rescales their dispersion to a common value, after \citet{rodrigues2019rpa} but without the rotation step; Section~\ref{sec:baselines} gives the method and the reason.

\begin{table}[tb]
\centering
\caption{Alignment baselines against no defense and the proposed transform,
cross-session S1$\to$S2, at each dataset's reported operating point and under
the attacker configuration of Section~\ref{sec:setup}. EA is
\citet{he2020ea}; RPA-RS is recentring and scaling after \citet{rodrigues2019rpa}, without its rotation step (Section~\ref{sec:baselines}). Decoder is FBCNet on BCI-IV-2a, EEGNet on OpenBMI.
Chance is \bciChance\ and \obmiChance. Every row uses the same attacker configuration, including XDAWN maximised
over $\mathrm{nfilter} \in \{1,2,4,8\}$; the RPA-RS rows select
$\mathrm{nfilter}=8$.}
\label{tab:app-ea}
\small
\setlength{\tabcolsep}{4pt}
\begin{tabular}{llccccccc}
\toprule
Dataset & Condition & SVM & XD & SPD & CSP-S & CSP-L & Worst & Decoder \\
\midrule
\multirow{4}{*}{BCI-IV-2a}
 & No defense & 90.8 & 87.6 & 89.9 & 73.2 & 77.3 & 90.8 & 71.9 \\
 & EA         & 60.3 & 69.0 & 23.0 & 48.6 & 29.5 & 69.0 & \textbf{74.8} \\
 & RPA-RS     & 46.3 & 34.0 & 25.1 & 39.5 & 22.5 & 46.3 & 69.8 \\
 & Ours       & 11.1 & 15.5 & 15.1 & 15.6 & \phantom{0}3.5 & \textbf{15.6} & 64.9 \\
\midrule
\multirow{4}{*}{OpenBMI}
 & No defense & 90.8 & 88.6 & 66.1 & 83.5 & 85.9 & 90.8 & 65.3 \\
 & EA         & 40.4 & 48.4 & 10.1 & 22.8 & 20.3 & 48.4 & \textbf{65.8} \\
 & RPA-RS     & 33.9 & 54.5 & \phantom{0}3.4 & 24.0 & 20.2 & 54.5 & 63.6 \\
 & Ours       & \phantom{0}2.4 & \phantom{0}9.9 & \phantom{0}1.5 & \phantom{0}2.8 & \phantom{0}3.8 & \textbf{\phantom{0}9.9} & 63.1 \\
\bottomrule
\end{tabular}
\end{table}

EA reduces every attacker on both datasets and improves decoding accuracy
above the undefended baseline, which is what a transfer normalisation is
designed to do. The suppression is uneven: SPDNet falls furthest (89.9\% to
23.0\% on BCI-IV-2a), while XDAWN falls least and binds the worst case on
both datasets. XDAWN reads an evoked-augmented covariance that per-subject
whitening does not fully reach.

The gap to a purpose-built transform is large. EA leaves 69.0\% and 48.4\%
where the reported transform leaves \bciWorst\ and \obmiWorst, against chance
of \bciChance\ and \obmiChance. A practitioner running EA for transfer
reasons therefore has partial and unquantified protection, which is a
different position from having none but not one a release policy could rest
on. RPA-RS suppresses further than EA on BCI-IV-2a, 46.3\% against 69.0\%, but
not on OpenBMI, where XDAWN reads 54.5\% against EA's 48.4\%. Both leave the
worst attacker at four and twenty-nine times chance respectively.

\section{Comparison with Perturbation-Based Protection}
\label{app:meng}

\citet{meng2023identity} add a learned user-wise perturbation $\delta$ to
each subject's trials so that a user-classifier trained on the perturbed
release fails to generalise. It is the closest prior work, and it targets a
different representation from ours, so we reproduce it and run each method
against the other's attackers.


\paragraph{Reproduction.} We fit $\delta$ following their procedure and
evaluate under their protocol on BCI-IV-2a: user-identification accuracy
(UIA) for three deep time-domain classifiers, on unperturbed and on
user-wise perturbed data, with balanced classification accuracy (BCA) as
the utility measure.

\begin{table}[tb]
\centering
\caption{Reproduction of \citet{meng2023identity}, BCI-IV-2a. UIA is user-identification accuracy, unperturbed and under their user-wise perturbation; Bal.\ Acc.\ is balanced motor-imagery accuracy on the perturbed data. Their reported figures are given for comparison.}
\label{tab:meng-repro}
\small
\setlength{\tabcolsep}{5pt}
\begin{tabular}{lcccccc}
\toprule
& \multicolumn{2}{c}{UIA unperturbed} & \multicolumn{2}{c}{UIA perturbed}
& \multicolumn{2}{c}{Bal. Acc.} \\
\cmidrule(lr){2-3}\cmidrule(lr){4-5}\cmidrule(lr){6-7}
Classifier & Ours & Theirs & Ours & Theirs & Ours & Theirs \\
\midrule
EEGNet       & 68.7 & 67.1 & \phantom{0}8.5 & 15.7 & 55.2 & 52.0 \\
DeepConvNet  & 66.2 & 72.8 & 15.6 & 48.6 & 47.2 & 53.6 \\
ShallowConvNet & 89.5 & 87.9 & 19.9 & 18.9 & 55.0 & 53.9 \\
\bottomrule
\end{tabular}
\end{table}

Unperturbed UIA reproduces closely: 68.7\% against their 67.1\% for EEGNet
and 89.5\% against 87.9\% for ShallowConvNet. Perturbed UIA agrees for ShallowConvNet and is lower than theirs for EEGNet and DeepConvNet, 8.5\% against 15.7\% and 15.6\% against 48.6\%; our reproduction therefore suppresses that classifier more
than they report, so the comparison below is not favourable to us by
construction.

\paragraph{Each method against the other's attackers.} The two defenses
address disjoint representations, and applying each to the other's threat
model makes that concrete.

\begin{table}[tb]
\centering
\caption{Cross-evaluation on BCI-IV-2a at the reported operating point
$(0.9,0.35,0.6,0.0)$, under the attacker configuration of
Section~\ref{sec:setup}. The perturbation of \citet{meng2023identity} leaves
every covariance-space attacker at its undefended value, and the proposed
transform reduces all five.}
\label{tab:meng-cross}
\small
\setlength{\tabcolsep}{5pt}
\begin{tabular}{lcccccc}
\toprule
Condition & SVM & XD & SPD & CSP-S & CSP-L & Worst \\
\midrule
No defense & 90.8 & 87.6 & 89.9 & 73.2 & 77.3 & 90.8 \\
\citet{meng2023identity} $\delta$ & 90.8 & 87.5 & 89.9 & 73.5 & 77.2 & 90.8 \\
Ours & 11.1 & 15.5 & 15.1 & 15.6 & \phantom{0}3.5 & \textbf{15.6} \\
\bottomrule
\end{tabular}
\end{table}

Their perturbation moves no covariance attacker by more than 0.3 points, and
the worst case is unchanged at \bciBase. Conversely,
Section~\ref{sec:boundary} shows our transform leaves deep time-domain identification at 83.3\% on BCI-IV-2a where theirs reduces it to 8.5--19.9\%.
Section~\ref{sec:baselines} discusses what follows.

\section{Reproduction of a Published Single-Session Attack}
\label{app:bakjeong}

The five attackers of Section~\ref{sec:setup} are constructed for this paper.
To check the defense against an attack designed independently of it, we
reproduce the EEG subject-identification method of
\citet{bak2023identification} on OpenBMI and evaluate it on defended data
under its own protocol.

\paragraph{Protocol.} \citet{bak2023identification} classify one user against
the remaining 53 using CSP features and an RBF SVM, under 10-fold
cross-validation repeated thirty times, and validate the result with the half
total error rate $\mathrm{HTER} = (\mathrm{FAR} + \mathrm{FRR})/2$, treating
$\mathrm{HTER} \ge 0.5$ as a failure of identification. Their preprocessing
band-passes to 8--30\,Hz and segments 0--4\,s of the imagery period, which
matches the band this paper operates on. CSP features are log-variances of
the $2m$ retained components (their Eq.~2); we use $m = 4$ with unweighted
classes, confirmed by a search over an 18-subject subset.

\paragraph{Montage.} They select 18 motor channels over the parietal region
and 18 non-motor channels over frontal and occipital regions, from the
62-channel recording. OpenBMI as used elsewhere in this paper retains 20
motor channels, so we evaluate both: our 20-channel montage, and a
36-channel reconstruction following their motor/non-motor split.

\begin{table}[tb]
\centering
\caption{Reproduction of \citet{bak2023identification} on OpenBMI, single-session, undefended and at the reported operating point
$(0.9,0.60,0.6,0.2)$. Binary accuracy is the same-versus-different decision;
HTER is the half total error rate, chance $0.5$.}
\label{tab:bakjeong}
\small
\setlength{\tabcolsep}{5pt}
\begin{tabular}{llcccc}
\toprule
Montage & Condition & Binary acc. & FAR & FRR & HTER \\
\midrule
\multirow{2}{*}{20-channel}
 & No defense & 99.78 & 0.0004 & 0.098 & 0.049 \\
 & Defended   & 98.16 & 0.0001 & 0.989 & \textbf{0.494} \\
\midrule
\multirow{2}{*}{36-channel}
 & No defense & 99.86 & 0.0003 & 0.059 & 0.030 \\
 & Defended   & 98.37 & 0.0004 & 0.859 & \textbf{0.430} \\
\bottomrule
\end{tabular}
\end{table}

The reproduction is close on their own headline quantity: averaged over the
54 users, our one-against-53 accuracy is 99.78\% (sd 0.22, range 99.09--100)
against the 98.97\% they report for CSP--SVM, at HTER 0.049 against their
0.01 $\pm$ 0.00. On defended data the HTER rises to 0.494 on the 20-channel
montage and 0.430 on the 36-channel one.

The interpretation is theirs. \citet{bak2023identification} apply the same
pipeline to Dataset~IVa, a five-subject dataset, and obtain 56.2\% accuracy
at HTER $0.5 \pm 0.00$; they treat that as the point at which CSP-based
identification ceases to be reliable, and it motivates their procedure for
estimating a minimum cohort size. Our defended 20-channel condition reaches 0.494 on 54 subjects, the same
failure threshold, and on a dataset they show is comfortably large enough for
the method to work.

The binary accuracy stays high because the decision is dominated by the
negative class: FAR falls to $10^{-4}$ while FRR rises to 0.99, so the
classifier accepts almost nothing and the same-subject pairs it is supposed
to accept are rejected. That is the failure mode a verification metric is
designed to expose and a raw accuracy would hide, which is why HTER rather
than accuracy is the quantity to read.

This attack operates within a single session and is therefore easier than the
cross-session protocol the rest of the paper uses
(Section~\ref{sec:setup}); we report it because it is a published method
evaluated on its own terms.

\section{Sub-band Operation}
\label{app:subband}

Stages 1--2 equalise the 8--30\,Hz covariance as one object, and
Section~\ref{sec:ablation} shows that a stage which re-aligns a narrower slice
too aggressively can re-introduce structure CSP reads. That raises the
question of whether an attacker gains by the transform being applied within a
narrow band rather than across the whole release. We therefore repeat the
defense with the transform fitted and applied inside each of three narrow
bands, adding the residual 8–30 Hz component back untouched, and read the released
signal with the same CSP attackers at the matching band. The release is
band-limited to 8--30\,Hz in every condition, so only the band the transform
operates on varies. ``Stacked'' is the strongest of the three narrow-band
attackers, the choice an adversary free to pick a band would make.

\begin{table}[tb]
\centering
\caption{Sub-band operation, cross-session S1$\to$S2 at each dataset's
reported operating point, under the attacker configuration of
Section~\ref{sec:setup}. The full-band row is the reported defense and
reproduces Table~\ref{tab:main}'s CSP cells. Chance is \bciChance\ and
\obmiChance.}
\label{tab:app-subband}
\small
\setlength{\tabcolsep}{6pt}
\begin{tabular}{lcc@{\hskip 14pt}cc}
\toprule
& \multicolumn{2}{c}{BCI-IV-2a $(0.9,0.35,0.6,0.0)$}
& \multicolumn{2}{c}{OpenBMI $(0.9,0.60,0.6,0.2)$} \\
\cmidrule(lr){2-3}\cmidrule(lr){4-5}
Band & CSP-SVM & CSP-LDA & CSP-SVM & CSP-LDA \\
\midrule
full 8--30\,Hz   & 15.6 & \phantom{0}3.5 & 2.8 & 3.8 \\
$\mu$ 8--12\,Hz  & 13.4 & \phantom{0}6.8 & 2.1 & 2.3 \\
lo-$\beta$ 12--20\,Hz & 12.8 & \phantom{0}4.9 & 2.3 & 3.9 \\
hi-$\beta$ 20--30\,Hz & 12.8 & \phantom{0}4.8 & 2.5 & 4.1 \\
\midrule
stacked          & 13.4 & \phantom{0}6.8 & 2.5 & 4.1 \\
\bottomrule
\end{tabular}
\end{table}

At the reported operating points the attacker does not gain. CSP-SVM, which
binds the worst case on BCI-IV-2a, is \emph{lower} in every narrow band than
at full band, 12.8--13.4 against 15.6, and on OpenBMI it falls from 2.8 to
2.1--2.5. CSP-LDA rises on BCI-IV-2a, from 3.5\% to at most 6.8\%, and is
flat on OpenBMI within two points. No band, and no choice among bands, lifts
either attacker above the full-band worst case of \bciWorst\ and \obmiWorst.

The mechanism Section~\ref{sec:ablation} identifies is therefore reachable
only by mis-setting a strength, not by the adversary's choice of band: it is
$\alpha = 1.0$ and $\beta = 1.0$ that let CSP recover
(Tables~\ref{tab:app-alpha} and \ref{tab:app-beta}), and both lie outside the operating-point selection grid of Section~\ref{sec:setup}.


\section{Session-Level Linkage}
\label{app:linkage}

Subject templates are built from defended Session~1; $N$ defended trials from
Session~2 are drawn at random and aggregated into a single decision. Subsets
are drawn only from Session~2, and neither the templates nor the classifier
sees a Session-2 label. Each batch size is repeated twenty times with a fresh
subset and we report mean and standard deviation; the ``all'' level uses
every trial a subject contributes and so has one deterministic value.

Four aggregation rules are used. \emph{Mean tangent} takes the nearest
Session-1 template to the mean tangent-space embedding of the batch.
\emph{Logit mean} takes the argmax of the mean classifier decision score.
\emph{Max logit} takes the argmax of the single most confident trial in the
batch, which is what an adversary unwilling to tolerate false links would
use. \emph{Majority} classifies each trial and takes the modal label.
Summing rather than averaging the decision scores is not a fifth rule:
$\sum_i z_i = N \bar{z}_i$ and $\arg\max$ is scale-invariant, so the two
coincide except where batch sizes differ.

\begin{table}[tb]
\centering
\caption{Session-level linkage on BCI-IV-2a at $(0.9,0.35,0.6,0.0)$, rank-1
accuracy (\%) over twenty draws. Chance is \bciChance. With nine subjects
accuracy is quantised in steps of 11.1 points, so a single subject changing
outcome moves a cell by that much; individual defended cells should not be
read as differing from one another.}
\label{tab:app-linkage-bci}
\small
\setlength{\tabcolsep}{4pt}
\begin{tabular}{lcccc}
\toprule
Batch $N$ & Mean tangent & Logit mean & Max logit & Majority \\
\midrule
\multicolumn{5}{l}{\emph{No defense}} \\
\phantom{0}1 & \phantom{0}88.3 $\pm$ 8.9 & 92.2 $\pm$ 5.1 & 92.2 $\pm$ 5.1 & 92.2 $\pm$ 5.1 \\
\phantom{0}5 & \phantom{0}97.2 $\pm$ 4.8 & 88.9 $\pm$ 0.0 & 89.4 $\pm$ 2.4 & 88.9 $\pm$ 0.0 \\
10 & \phantom{0}96.7 $\pm$ 5.1 & 88.9 $\pm$ 0.0 & 88.9 $\pm$ 0.0 & 89.4 $\pm$ 2.4 \\
25 & \phantom{0}98.9 $\pm$ 3.3 & 88.9 $\pm$ 0.0 & 88.9 $\pm$ 0.0 & 88.9 $\pm$ 0.0 \\
50 & 100.0 $\pm$ 0.0 & 88.9 $\pm$ 0.0 & 88.9 $\pm$ 0.0 & 88.9 $\pm$ 0.0 \\
all & 100.0 $\pm$ 0.0 & 88.9 $\pm$ 0.0 & 88.9 $\pm$ 0.0 & 88.9 $\pm$ 0.0 \\
\midrule
\multicolumn{5}{l}{\emph{Defended}} \\
\phantom{0}1 & \phantom{00}3.9 $\pm$ 5.3 & 11.1 $\pm$ 0.0 & 11.1 $\pm$ 0.0 & 11.1 $\pm$ 0.0 \\
\phantom{0}5 & \phantom{00}2.2 $\pm$ 4.4 & 11.1 $\pm$ 0.0 & 11.1 $\pm$ 0.0 & 11.1 $\pm$ 0.0 \\
10 & \phantom{00}1.1 $\pm$ 3.3 & 11.1 $\pm$ 0.0 & 11.1 $\pm$ 0.0 & 11.1 $\pm$ 0.0 \\
25 & \phantom{00}1.1 $\pm$ 3.3 & 11.1 $\pm$ 0.0 & 11.1 $\pm$ 0.0 & 11.1 $\pm$ 0.0 \\
50 & \phantom{00}1.1 $\pm$ 3.3 & 11.1 $\pm$ 0.0 & 11.1 $\pm$ 0.0 & 11.1 $\pm$ 0.0 \\
all & \phantom{00}0.0 $\pm$ 0.0 & 11.1 $\pm$ 0.0 & 11.1 $\pm$ 0.0 & 11.1 $\pm$ 0.0 \\
\bottomrule
\end{tabular}
\end{table}

\begin{table}[tb]
\centering
\caption{Session-level linkage on OpenBMI at $(0.9,0.60,0.6,0.2)$, rank-1
accuracy (\%) over twenty draws. Chance is \obmiChance. With 54 subjects the
quantisation is 1.85 points, so these cells resolve the claim that the
BCI-IV-2a table cannot.}
\label{tab:app-linkage-ob}
\small
\setlength{\tabcolsep}{4pt}
\begin{tabular}{lcccc}
\toprule
Batch $N$ & Mean tangent & Logit mean & Max logit & Majority \\
\midrule
\multicolumn{5}{l}{\emph{No defense}} \\
\phantom{0}1 & 82.5 $\pm$ 2.6 & 90.3 $\pm$ 2.0 & 90.3 $\pm$ 2.0 & 90.4 $\pm$ 2.1 \\
\phantom{0}5 & 87.5 $\pm$ 2.0 & 90.6 $\pm$ 2.1 & 90.2 $\pm$ 1.8 & 92.3 $\pm$ 1.8 \\
10 & 89.4 $\pm$ 1.8 & 91.7 $\pm$ 1.1 & 90.1 $\pm$ 1.3 & 93.3 $\pm$ 1.4 \\
25 & 89.8 $\pm$ 1.1 & 92.2 $\pm$ 0.9 & 90.5 $\pm$ 1.6 & 93.5 $\pm$ 0.9 \\
50 & 90.3 $\pm$ 0.8 & 92.6 $\pm$ 0.0 & 90.0 $\pm$ 1.2 & 93.4 $\pm$ 1.1 \\
all & 90.7 $\pm$ 0.0 & 92.6 $\pm$ 0.0 & 90.7 $\pm$ 0.0 & 94.4 $\pm$ 0.0 \\
\midrule
\multicolumn{5}{l}{\emph{Defended}} \\
\phantom{0}1 & \phantom{0}1.6 $\pm$ 1.6 & \phantom{0}2.6 $\pm$ 1.1 & \phantom{0}2.6 $\pm$ 1.1 & \phantom{0}2.6 $\pm$ 1.1 \\
\phantom{0}5 & \phantom{0}0.6 $\pm$ 1.1 & \phantom{0}2.0 $\pm$ 0.6 & \phantom{0}2.4 $\pm$ 1.0 & \phantom{0}2.2 $\pm$ 0.7 \\
10 & \phantom{0}1.3 $\pm$ 0.8 & \phantom{0}2.1 $\pm$ 0.7 & \phantom{0}2.5 $\pm$ 1.1 & \phantom{0}2.0 $\pm$ 0.6 \\
25 & \phantom{0}0.4 $\pm$ 0.7 & \phantom{0}1.9 $\pm$ 0.0 & \phantom{0}2.4 $\pm$ 1.0 & \phantom{0}1.9 $\pm$ 0.0 \\
50 & \phantom{0}0.5 $\pm$ 0.8 & \phantom{0}1.9 $\pm$ 0.0 & \phantom{0}2.5 $\pm$ 1.1 & \phantom{0}1.9 $\pm$ 0.0 \\
all & \phantom{0}1.9 $\pm$ 0.0 & \phantom{0}1.9 $\pm$ 0.0 & \phantom{0}3.7 $\pm$ 0.0 & \phantom{0}1.9 $\pm$ 0.0 \\
\bottomrule
\end{tabular}
\end{table}

\begin{figure}[tb]
\centering
\includegraphics[width=1\textwidth]{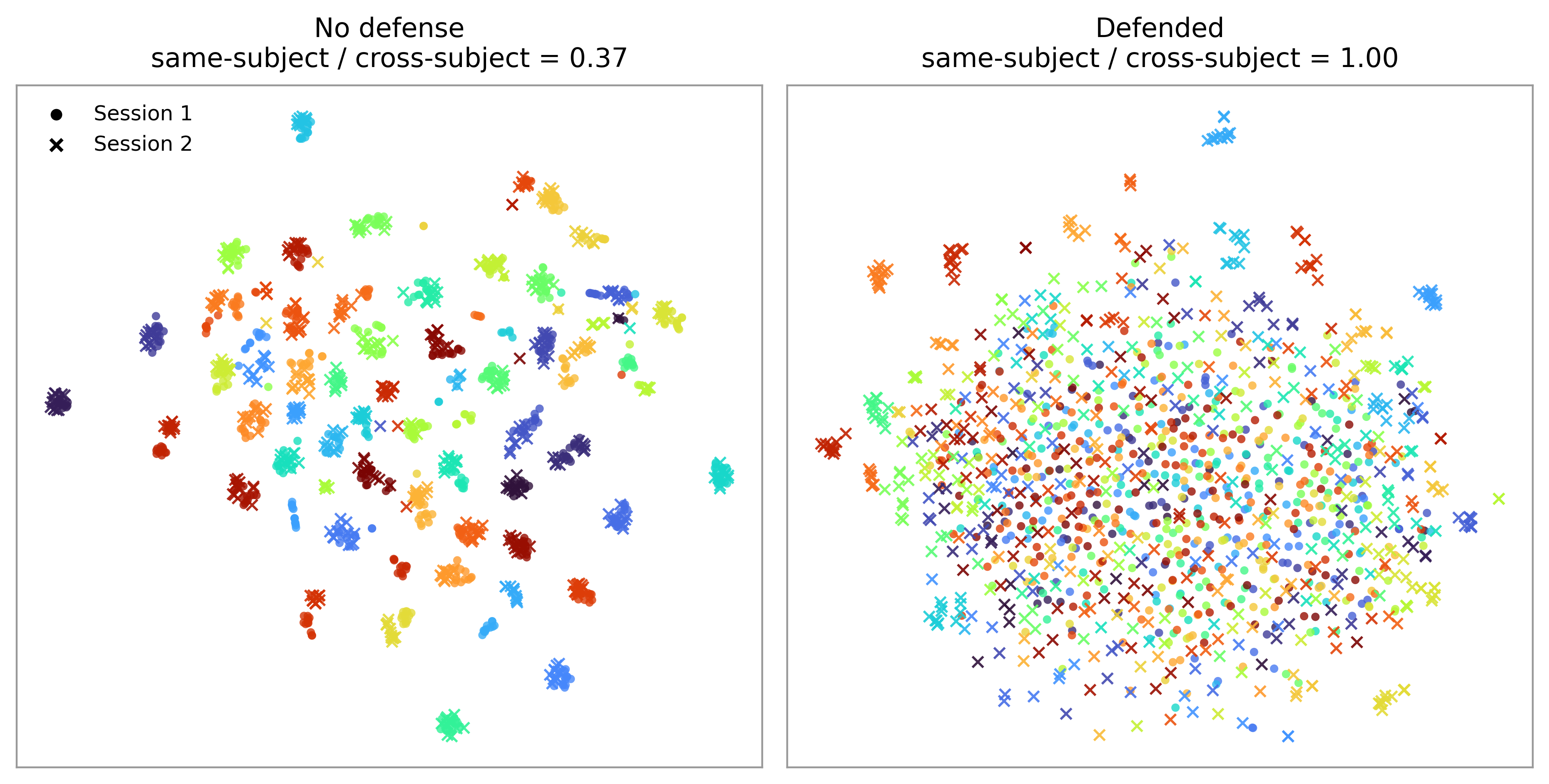}
\caption{Cross-session linkage on OpenBMI, Session~1 (circles) against
Session~2 (crosses), coloured by subject, t-SNE from the tangent space at the
population Riemannian mean. Undefended, each subject's sessions form one
cluster; defended, they do not. The ratio above each panel is computed in the
full 210 dimensions, not the projection, and t-SNE is fitted per panel so
distances are not comparable between them.}
\label{fig:link}
\end{figure}

Undefended, aggregation helps the adversary, which is the concern that
motivates measuring linkage separately from trial-level identification. The
mean-tangent rule gains most because averaging suppresses per-trial noise
while leaving the subject's mean covariance intact: it rises from 88.3\% to
100\% on BCI-IV-2a and from 82.5\% to 90.7\% on OpenBMI. The score-based
rules do not gain, because a per-trial classifier already at 90\% has little
left to recover. Figure~\ref{fig:link} shows the embedding.

Defended, no rule and no batch size recovers identity. The mean-tangent rule
in fact \emph{declines} as trials are added, 3.9\% to 0.0\% on BCI-IV-2a and
1.6\% to 0.4\% at $N{=}25$ on OpenBMI. What the transform leaves behind is
variation around the shared population mean rather than around a subject's
own, so averaging drives the batch toward the population and away from any
template. Max logit is the only rule that rises at the ``all'' level (3.7\%
on OpenBMI, twice chance), which is expected: taking the single most
confident trial out of two hundred selects the tail of a distribution rather
than its centre, and is the rule least helped by the defense.

\section{Identity and Task, Measured the Same Way}
\label{app:sep}

For each channel we compute the squared deviation of a group mean log band
power from its reference, in units of the relevant within-group trial
variance: for subjects, a subject's mean against the population, in units of
that subject's own trial variance; for classes, a class mean against that
subject's own mean, in units of within-class variance, computed inside each
subject and then averaged, since a decoder is fitted per subject and never
compares band power across people. Under the transform every class retains
between 51\% and 58\% of its deviation on BCI-IV-2a. The two datasets differ
in cohort size, montage and class count, so the ordering is not an artefact
of either.

\section{Parameter Sweeps}
\label{app:sensitivity}

This section sweeps each of the four strengths in turn, with the other three held at that dataset's reported setting, cross-session S1$\to$S2 on both datasets. The swept values extend beyond the selection grid of Section~\ref{sec:setup}, since these are sensitivity analyses rather than part of the selection. The $\gamma_{\mathrm{rma}}$ and $\gamma_{\mathrm{scw}}$ columns for
BCI-IV-2a are Figure~\ref{fig:opcurve} in tabular form; the rest are new.
They are tables rather than figures because $\alpha$ and $\beta$ span too
little range to plot usefully, which is itself the result.

These sweeps report four of the five attackers; SPDNet-ID does not bind at
either reported operating point and is omitted so the tables show the
attackers that determine the worst case. The four-attacker worst case at both
reported points therefore coincides with the five-attacker worst of
Table~\ref{tab:main}.

\begin{table}[tb]
\centering
\caption{Subject-mean alignment strength $\gamma_{\mathrm{rma}}$,
cross-session, with $(\gamma_{\mathrm{scw}},\alpha,\beta)$ held at
$(0.35,0.6,0.0)$ on BCI-IV-2a and $(0.60,0.6,0.2)$ on OpenBMI. Worst is over
SVM, XDAWN, CSP-SVM and CSP-LDA. Utility is FBCNet on BCI-IV-2a, EEGNet on
OpenBMI. Chance is \bciChance\ and \obmiChance.}
\label{tab:app-grma}
\small
\setlength{\tabcolsep}{5pt}
\begin{tabular}{lcccccc@{\hskip 12pt}cccccc}
\toprule
& \multicolumn{6}{c}{BCI-IV-2a} & \multicolumn{6}{c}{OpenBMI} \\
\cmidrule(lr){2-7}\cmidrule(lr){8-13}
$\gamma_{\mathrm{rma}}$ & SVM & XD & CSP-S & CSP-L & Worst & FBCNet
       & SVM & XD & CSP-S & CSP-L & Worst & EEGNet \\
\midrule
0.0 & 53.5 & 50.9 & 43.0 & 31.1 & 53.5 & 51.2
    & 60.4 & 54.1 & 50.7 & 61.4 & 61.4 & 56.8 \\
0.2 & 34.5 & 33.5 & 19.3 & 10.7 & 34.5 & 56.3
    & 41.1 & 35.9 & 33.0 & 44.4 & 44.4 & 58.9 \\
0.4 & 19.5 & 23.1 & 11.8 & \phantom{0}8.1 & 23.1 & 59.9
    & 21.1 & 23.4 & 16.6 & 21.4 & 23.4 & 60.3 \\
0.6 & \phantom{0}9.9 & 16.9 & 10.5 & \phantom{0}5.7 & 16.9 & 63.1
    & \phantom{0}7.4 & 14.6 & \phantom{0}7.1 & 12.6 & 14.6 & 61.8 \\
0.8 & 11.2 & 15.1 & 14.5 & \phantom{0}5.0 & 15.1 & 64.1
    & \phantom{0}2.8 & 10.4 & \phantom{0}3.3 & \phantom{0}5.6 & 10.4 & 62.4 \\
0.9 & 11.1 & 15.5 & 15.6 & \phantom{0}3.5 & \textbf{15.6} & \textbf{64.9}
    & \phantom{0}2.4 & \phantom{0}9.9 & \phantom{0}2.8 & \phantom{0}3.8 & \textbf{\phantom{0}9.9} & \textbf{63.1} \\
1.0 & 11.1 & 15.7 & 15.3 & 14.5 & 15.7 & 64.7
    & \phantom{0}2.8 & 10.6 & \phantom{0}3.2 & \phantom{0}3.9 & 10.6 & 62.9 \\
\bottomrule
\end{tabular}
\end{table}

Both datasets behave identically on this axis: raising $\gamma_{\mathrm{rma}}$
from 0 to the selected $0.9$ improves the worst attacker from 53.5\% to
\bciWorst\ and from 61.4\% to \obmiWorst\ \emph{and} the decoder from 51.2\%
to 64.9\% and from 56.8\% to 63.1\%, with $1.0$ slightly worse on both axes
in each case. Alignment is not a trade, which is what
Section~\ref{sec:ablation}'s claim about RMA rests on.

\begin{table}[tb]
\centering
\caption{Whitening strength $\gamma_{\mathrm{scw}}$, cross-session, with
$(\gamma_{\mathrm{rma}},\alpha,\beta)$ held at $(0.9,0.6,0.0)$ on BCI-IV-2a
and $(0.9,0.6,0.2)$ on OpenBMI. Columns as in Table~\ref{tab:app-grma}. Chance is \bciChance\ and \obmiChance.}
\label{tab:app-gscw}
\small
\setlength{\tabcolsep}{5pt}
\begin{tabular}{lcccccc@{\hskip 12pt}cccccc}
\toprule
& \multicolumn{6}{c}{BCI-IV-2a} & \multicolumn{6}{c}{OpenBMI} \\
\cmidrule(lr){2-7}\cmidrule(lr){8-13}
$\gamma_{\mathrm{scw}}$ & SVM & XD & CSP-S & CSP-L & Worst & FBCNet
        & SVM & XD & CSP-S & CSP-L & Worst & EEGNet \\
\midrule
0.20 & 11.5 & 19.1 & 20.2 & \phantom{0}3.4 & 20.2 & \textbf{68.9}
     & 10.3 & 37.5 & 11.4 & 12.7 & 37.5 & \textbf{64.7} \\
0.35 & 11.1 & 15.5 & 15.6 & \phantom{0}3.5 & \textbf{15.6} & 64.9
     & \phantom{0}5.2 & 21.4 & \phantom{0}6.2 & \phantom{0}8.1 & 21.4 & 63.9 \\
0.45 & 11.1 & 13.4 & 14.1 & \phantom{0}4.1 & 14.1 & 62.9
     & \phantom{0}4.1 & 15.7 & \phantom{0}4.7 & \phantom{0}6.1 & 15.7 & 63.3 \\
0.50 & 11.1 & 12.7 & 13.5 & \phantom{0}4.2 & 13.5 & 61.3
     & \phantom{0}3.3 & 13.1 & \phantom{0}4.0 & \phantom{0}5.6 & 13.1 & 63.5 \\
0.55 & 11.1 & 11.9 & 13.2 & \phantom{0}4.1 & 13.2 & 60.2
     & \phantom{0}2.9 & 11.3 & \phantom{0}3.4 & \phantom{0}4.6 & 11.3 & 63.1 \\
0.60 & 11.1 & 11.5 & 13.0 & \phantom{0}4.9 & 13.0 & 58.1
     & \phantom{0}2.4 & \phantom{0}9.9 & \phantom{0}2.8 & \phantom{0}3.8 & \textbf{\phantom{0}9.9} & 63.0 \\
0.70 & 11.1 & 11.6 & 12.8 & \phantom{0}3.7 & \textbf{12.8} & 55.2
     & \phantom{0}2.0 & \phantom{0}8.5 & \phantom{0}2.1 & \phantom{0}3.1 & \phantom{0}8.5 & 62.3 \\
0.80 & 11.1 & 11.6 & 13.2 & \phantom{0}3.4 & 13.2 & 53.2
     & \phantom{0}1.5 & \phantom{0}7.5 & \phantom{0}2.3 & \phantom{0}1.9 & \textbf{\phantom{0}7.5} & 61.9 \\
\bottomrule
\end{tabular}
\end{table}

$\gamma_{\mathrm{scw}}$, by contrast, is where the trade lives: on BCI-IV-2a
the worst attacker falls 20.2\% to 12.8\% while FBCNet falls 68.9\% to
53.2\%, roughly a point of accuracy per point of privacy. OpenBMI is more
favourable: the worst attacker falls 37.5\% to 7.5\% for 2.9 points of
EEGNet. That is why the selected $\gamma_{\mathrm{scw}}$ is 0.60 there
against 0.35 on BCI-IV-2a, and why the whitening strength is the one
parameter that must be chosen per cohort.

\begin{table}[tb]
\centering
\caption{Band-Centre Alignment strength $\alpha$, cross-session, with
$(\gamma_{\mathrm{rma}},\gamma_{\mathrm{scw}},\beta)$ held at
$(0.9,0.35,0.0)$ on BCI-IV-2a and $(0.9,0.60,0.2)$ on OpenBMI. Worst is over
SVM, XDAWN, CSP-SVM and CSP-LDA. Utility is FBCNet on BCI-IV-2a, EEGNet on
OpenBMI. Chance is \bciChance\ and \obmiChance.}
\label{tab:app-alpha}
\small
\setlength{\tabcolsep}{5pt}
\begin{tabular}{lcccccc@{\hskip 12pt}cccccc}
\toprule
& \multicolumn{6}{c}{BCI-IV-2a} & \multicolumn{6}{c}{OpenBMI} \\
\cmidrule(lr){2-7}\cmidrule(lr){8-13}
$\alpha$ & SVM & XD & CSP-S & CSP-L & Worst & FBCNet
         & SVM & XD & CSP-S & CSP-L & Worst & EEGNet \\
\midrule
0.0 & 11.3 & 15.3 & 13.3 & \phantom{0}5.7 & 15.3 & \textbf{65.3}
    & \phantom{0}2.9 & 10.5 & 3.4 & 5.4 & 10.5 & 63.3 \\
0.2 & 11.2 & 15.4 & 14.3 & \phantom{0}4.9 & 15.4 & 65.1
    & \phantom{0}2.7 & 10.3 & 3.2 & 4.8 & 10.3 & \textbf{63.6} \\
0.4 & 11.1 & 15.2 & 14.7 & \phantom{0}4.4 & \textbf{15.2} & 64.8
    & \phantom{0}2.4 & 10.2 & 2.9 & 4.4 & 10.2 & 63.3 \\
0.6 & 11.1 & 15.5 & 15.6 & \phantom{0}3.5 & 15.6 & 64.9
    & \phantom{0}2.4 & \phantom{0}9.9 & 2.8 & 3.8 & \textbf{\phantom{0}9.9} & 63.1 \\
0.8 & 11.1 & 15.4 & 15.0 & \phantom{0}2.9 & 15.4 & 64.7
    & \phantom{0}2.5 & 10.2 & 3.2 & 4.1 & 10.2 & 63.3 \\
1.0 & 11.1 & 15.9 & 16.2 & 13.7 & 16.2 & 64.8
    & \phantom{0}2.7 & 10.4 & 3.0 & 4.4 & 10.4 & 62.9 \\
\bottomrule
\end{tabular}
\end{table}

On OpenBMI the $\alpha$ curve is shallow but has its minimum exactly at the
selected $0.6$: the worst attacker spans 9.9--10.5 and EEGNet 62.9--63.6, so
the Session-1 procedure chose the cross-session optimum without seeing
Session~2. On BCI-IV-2a it does not: the worst attacker spans 15.2--15.6 for
$\alpha \le 0.8$ with the minimum at $0.4$ rather than the selected $0.6$, a
0.4-point difference inside run-to-run variation. Only $\alpha = 1.0$ is
clearly worse, where CSP-LDA rises from 2.9\% to 13.7\%.

\begin{table}[tb]
\centering
\caption{Cleanup strength $\beta$, cross-session, with
$(\gamma_{\mathrm{rma}},\gamma_{\mathrm{scw}},\alpha)$ held at
$(0.9,0.35,0.6)$ on BCI-IV-2a and $(0.9,0.60,0.6)$ on OpenBMI. Columns as in
Table~\ref{tab:app-alpha}.}
\label{tab:app-beta}
\small
\setlength{\tabcolsep}{5pt}
\begin{tabular}{lcccccc@{\hskip 12pt}cccccc}
\toprule
& \multicolumn{6}{c}{BCI-IV-2a} & \multicolumn{6}{c}{OpenBMI} \\
\cmidrule(lr){2-7}\cmidrule(lr){8-13}
$\beta$ & SVM & XD & CSP-S & CSP-L & Worst & FBCNet
        & SVM & XD & CSP-S & CSP-L & Worst & EEGNet \\
\midrule
0.0 & 11.1 & 15.5 & 15.6 & \phantom{0}3.5 & \textbf{15.6} & 64.9
    & 2.4 & 10.1 & 2.9 & 4.1 & 10.1 & 63.1 \\
0.1 & 11.1 & 15.6 & 15.9 & \phantom{0}3.3 & 15.9 & 64.6
    & 2.4 & 10.1 & 2.8 & 4.0 & 10.1 & \textbf{63.2} \\
0.2 & 11.1 & 15.6 & 16.0 & \phantom{0}2.8 & 16.0 & 64.7
    & 2.4 & \phantom{0}9.9 & 2.8 & 3.8 & \textbf{\phantom{0}9.9} & 63.1 \\
0.3 & 11.1 & 15.3 & 16.4 & \phantom{0}2.4 & 16.4 & 64.6
    & 2.4 & \phantom{0}9.9 & 2.9 & 4.1 & \textbf{\phantom{0}9.9} & 63.0 \\
0.4 & 11.1 & 15.4 & 16.1 & \phantom{0}2.3 & 16.1 & 64.5
    & 2.4 & 10.0 & 2.9 & 4.1 & 10.0 & 62.9 \\
0.5 & 11.1 & 15.6 & 14.9 & \phantom{0}2.4 & \textbf{15.6} & 65.1
    & 2.4 & 10.1 & 3.0 & 3.9 & 10.1 & 63.1 \\
0.6 & 11.1 & 15.7 & 14.7 & \phantom{0}3.2 & 15.7 & 64.9
    & 2.5 & 10.1 & 3.1 & 4.6 & 10.1 & 62.9 \\
0.7 & 11.1 & 15.7 & 15.1 & \phantom{0}4.2 & 15.7 & 65.0
    & 2.5 & 10.2 & 3.0 & 4.5 & 10.2 & \textbf{63.2} \\
0.8 & 11.1 & 15.8 & 15.6 & \phantom{0}6.1 & 15.8 & \textbf{65.3}
    & 2.6 & 10.3 & 3.0 & 4.8 & 10.3 & 63.0 \\
0.9 & 11.1 & 15.9 & 15.2 & 12.2 & 15.9 & 65.1
    & 2.6 & 10.4 & 3.1 & 5.3 & 10.4 & 63.0 \\
1.0 & 11.1 & 15.9 & 14.9 & 18.7 & 18.7 & 64.9
    & 2.6 & 10.6 & 2.9 & 5.2 & 10.6 & 63.0 \\
\bottomrule
\end{tabular}
\end{table}

The cleanup stage behaves the same way on both datasets: the selected value
is at or beside the minimum, and the whole range is narrow. On OpenBMI the
worst attacker spans 9.9--10.6 with its minimum at the selected $0.2$ (tied
with $0.3$), and EEGNet spans 62.9--63.2. On BCI-IV-2a the worst spans
15.6--16.4 for $\beta \le 0.9$, with the minimum at the selected $\beta = 0$
tied with $0.5$, and FBCNet spans 64.5--65.3.

Both curves turn upward at full strength, and for the same reason: CSP-LDA,
which the earlier stages leave far below chance, rises sharply once the final
alignment is applied at full strength, from 2.4\% at $\beta = 0.5$ to
18.7\% at $\beta = 1.0$ on BCI-IV-2a, where it becomes the binding attacker.
A stage that re-equalises the broadband mean too aggressively re-introduces
narrow-band structure that CSP reads. This is the same failure the
$\alpha = 1.0$ column of Table~\ref{tab:app-alpha} shows, and it is why both
strengths are useful only at moderate values.

A $4^4$ cross-session sweep over all 256 configurations bounds what the
Session-1 procedure gives up. The lowest worst-case attacker anywhere in the
grid is 11.3\% at $(0.6,0.9,0.3,0.6)$, essentially the \bciChance\ chance
floor, but FBCNet there reads 45.2\%. That is twenty points below the
reported point and barely above the 25\% four-class chance level. Restricting to the
80 cells within half a point of the reported utility, the best worst-case is
15.2\% at $(0.9,0.35,0.0,0.4)$, against \bciWorst\ for the point selected
without seeing Session~2. Blind selection therefore costs 0.4 points of
privacy at matched utility, and the configurations that do better on the
privacy axis do so by abandoning the task.

\section{Full Statistical Significance Tables}
\label{app:stats}

Paired Wilcoxon signed-rank, Cohen's $d_z$, and a bootstrap 95\% confidence
interval, band-limited baseline versus defended.

\begin{table}[tb]
\centering
\caption{BCI-IV-2a HO, $n{=}9$ subjects. Positive differences denote a
reduction in attacker accuracy (privacy rows) or in decoding accuracy
(utility rows).}
\label{tab:app-stats-bciho}
\small
\setlength{\tabcolsep}{4pt}
\begin{tabular}{lcccccc}
\toprule
Metric & Base & Defended & Diff & 95\% CI & $d_z$ & Wilcoxon $p$ \\
\midrule
\multicolumn{7}{l}{\emph{Privacy}} \\
SVM      & 90.8 & 11.1 & $+79.7$ & [53.1, 99.8] & 2.00 & 0.0078 \\
XDAWN    & 87.6 & 15.5 & $+72.1$ & [47.4, 92.4] & 1.95 & 0.0078 \\
SPDNet   & 89.9 & 15.1 & $+74.8$ & [66.1, 83.8] & 5.21 & 0.0039 \\
CSP-SVM  & 73.2 & 15.6 & $+57.6$ & [32.6, 79.7] & 1.48 & 0.0156 \\
CSP-LDA  & 77.3 & \phantom{0}3.5 & $+73.8$ & [53.6, 87.2] & 2.55 & 0.0078 \\
\midrule
\multicolumn{7}{l}{\emph{Utility}} \\
FBCSP    & 58.0 & 49.0 & $+9.0$ & [2.9, 16.0] & 0.85 & 0.0391 \\
EEGNet   & 60.7 & 57.6 & $+3.1$ & [0.1, 7.9] & 0.46 & 0.1641 \\
DCN      & 55.7 & 52.8 & $+2.9$ & [$-0.1$, 7.1] & 0.48 & 0.1953 \\
FBCNet   & 71.9 & 64.6 & $+7.2$ & [3.7, 12.1] & 1.02 & 0.0039 \\
\bottomrule
\end{tabular}
\end{table}

\begin{table}[tb]
\centering
\caption{OpenBMI HO, $n{=}54$ subjects.}
\label{tab:app-stats-obho}
\small
\setlength{\tabcolsep}{4pt}
\begin{tabular}{lcccccc}
\toprule
Metric & Base & Defended & Diff & 95\% CI & $d_z$ & Wilcoxon $p$ \\
\midrule
\multicolumn{7}{l}{\emph{Privacy}} \\
SVM      & 90.8 & 2.4 & $+88.4$ & [81.3, 94.6] & 3.48 & $<$0.0001 \\
XDAWN    & 88.6 & 9.9 & $+78.7$ & [70.6, 85.9] & 2.73 & $<$0.0001 \\
SPDNet   & 66.1 & 1.5 & $+64.6$ & [56.8, 72.0] & 2.24 & $<$0.0001 \\
CSP-SVM  & 83.5 & 2.8 & $+80.7$ & [73.3, 87.3] & 3.03 & $<$0.0001 \\
CSP-LDA  & 85.9 & 3.8 & $+82.1$ & [74.1, 89.1] & 2.91 & $<$0.0001 \\
\midrule
\multicolumn{7}{l}{\emph{Utility}} \\
FBCSP    & 65.6 & 60.1 & $+5.5$ & [3.0, 8.2] & 0.56 & 0.0002 \\
EEGNet   & 65.3 & 63.1 & $+2.2$ & [0.9, 3.7] & 0.41 & 0.0060 \\
DCN      & 62.8 & 61.0 & $+1.8$ & [0.2, 3.3] & 0.30 & 0.0350 \\
FBCNet   & 64.9 & 57.9 & $+7.0$ & [4.0, 10.5] & 0.56 & $<$0.0001 \\
\bottomrule
\end{tabular}
\end{table}

All five attackers change significantly on both datasets. The utility
decoders do not: EEGNet and DeepConvNet on BCI-IV-2a fall short of $p<0.05$.
That asymmetry is the substantive point, with privacy gains of 57.6--88.4
percentage points against utility costs of 1.8--9.0.

\begin{table}[tb]
\centering
\caption{Event-related desynchronisation, $\beta$ band (13--30\,Hz),
BCI-IV-2a Session~2, mean over nine subjects with the between-subject
standard deviation. Negative values indicate desynchronisation. The interval
is a bootstrap 95\% CI on the per-subject raw-minus-defended difference;
intervals excluding zero are marked $^{*}$.}
\label{tab:erds}
\small
\setlength{\tabcolsep}{5pt}
\begin{tabular}{llccrl}
\toprule
Class & Channel & Raw ERDS & Defended ERDS & Diff & 95\% CI \\
\midrule
Right hand & C3 & $-12.02 \pm 12.8$ & $-7.22 \pm 16.9$ & $-4.80$ & $[-10.43, -0.82]^{*}$ \\
Right hand & Cz & $-11.33 \pm 11.1$ & $-9.40 \pm 9.5$ & $-1.93$ & $[\phantom{0}-3.42, -0.54]^{*}$ \\
Right hand & C4 & $-8.34 \pm 7.7$ & $-7.13 \pm 7.5$ & $-1.21$ & $[\phantom{0}-2.36, -0.12]^{*}$ \\
Left hand  & C3 & $-11.02 \pm 18.6$ & $-10.55 \pm 17.2$ & $-0.46$ & $[\phantom{0}-2.58, \phantom{0}1.64]$ \\
Left hand  & Cz & $-10.22 \pm 15.5$ & $-14.62 \pm 17.3$ & $+4.40$ & $[\phantom{0}-1.48, 14.10]$ \\
Left hand  & C4 & $-11.27 \pm 15.3$ & $-14.47 \pm 15.9$ & $+3.20$ & $[\phantom{0}-2.80, 12.54]$ \\
\bottomrule
\end{tabular}
\end{table}

\begin{table*}[tb]
\centering
\caption{Differential privacy on the trace-normalised trial covariance,
BCI-IV-2a cross-session, under the adjacency and sensitivity stated above. Attackers use the configuration of
Section~\ref{sec:setup}, including XDAWN maximised over
$\mathrm{nfilter} \in \{1,2,4,8\}$. Decoders average three seeds; the Ours row is Table~\ref{tab:main}'s. Chance is \bciChance\ and 25\%.}
\label{tab:dp}
\scriptsize
\setlength{\tabcolsep}{3.5pt}
\begin{tabular}{lcccccc@{\hskip 8pt}cccc}
\toprule
& \multicolumn{6}{c}{Re-identification $\downarrow$}
& \multicolumn{4}{c}{MI decoding $\uparrow$} \\
\cmidrule(lr){2-7}\cmidrule(lr){8-11}
Condition & SVM & XD & SPD & CSP-S & CSP-L & Worst
 & FBCSP & EEGNet & DCN & FBCNet \\
\midrule
No defense & 90.8 & 87.6 & 89.9 & 73.2 & 77.3 & 90.8
 & 58.0 & 61.2 & 55.8 & 71.9 \\
\midrule
\multicolumn{11}{l}{\emph{Central mode}} \\
Laplace, $\varepsilon{=}1$  & 11.1 & 16.5 & 22.2 & 21.7 & \phantom{0}0.0 & 22.2 & --- & 32.7 & 29.2 & 31.7 \\
Laplace, $\varepsilon{=}5$  & 11.1 & 16.3 & 15.2 & 21.7 & \phantom{0}0.0 & 21.7 & --- & 32.8 & 29.3 & 31.8 \\
Laplace, $\varepsilon{=}10$ & 11.1 & 16.0 & 12.1 & 21.8 & \phantom{0}0.0 & 21.8 & --- & 33.1 & 29.4 & 31.4 \\
Gaussian, $\varepsilon{=}1$  & 11.1 & 20.9 & 11.1 & 12.5 & 34.2 & 34.2 & --- & 33.8 & 30.3 & 32.3 \\
Gaussian, $\varepsilon{=}5$  & 11.1 & 19.4 & 11.1 & 12.5 & 29.2 & 29.2 & --- & 32.9 & 30.1 & 32.2 \\
Gaussian, $\varepsilon{=}10$ & 11.1 & 20.4 & 11.1 & 12.6 & 32.9 & 32.9 & --- & 33.3 & 29.8 & 31.9 \\
\midrule
\multicolumn{11}{l}{\emph{Local mode}} \\
Laplace, $\varepsilon{=}1$  & 25.5 & 28.7 & 17.9 & 26.4 & 24.9 & 28.7 & 32.0 & 36.0 & 32.9 & 35.3 \\
Laplace, $\varepsilon{=}5$  & 25.3 & 28.4 & 10.9 & 26.6 & 25.0 & 28.4 & 30.6 & 35.8 & 32.8 & 35.6 \\
Laplace, $\varepsilon{=}10$ & 25.3 & 28.7 & 11.1 & 26.9 & 24.8 & 28.7 & 30.3 & 35.6 & 33.1 & 35.0 \\
Gaussian, $\varepsilon{=}1$  & 25.2 & 29.6 & 23.5 & 26.1 & 25.7 & 29.6 & 30.2 & 36.0 & 33.2 & 34.7 \\
Gaussian, $\varepsilon{=}5$  & 25.1 & 29.1 & 19.5 & 25.5 & 26.0 & 29.1 & 31.9 & 36.0 & 32.3 & 34.5 \\
Gaussian, $\varepsilon{=}10$ & 25.5 & 29.2 & 23.3 & 25.6 & 25.9 & 29.2 & 30.6 & 36.2 & 32.7 & 34.9 \\
\midrule
\textbf{Ours} $(0.9,0.35,0.6,0.0)$
 & \textbf{11.1} & \textbf{15.5} & \textbf{15.1} & \textbf{15.6} & \phantom{0}\textbf{3.5} & \textbf{15.6}
 & \textbf{49.0} & \textbf{57.8} & \textbf{52.3} & \textbf{64.9} \\
\bottomrule
\end{tabular}
\end{table*}

\section{Effect on Task-Relevant Physiology}
\label{app:erds}

ERDS is computed by the band-power method of
\citet{pfurtscheller1999erd}: each trial is band-pass filtered to 13--30\,Hz,
every sample squared to give instantaneous power, the power averaged over
trials, and the task period expressed relative to a reference period as
$100 (P_{\mathrm{task}} - P_{\mathrm{ref}}) / P_{\mathrm{ref}}$, so negative
values denote desynchronisation. The value is computed per subject and the
nine subject values averaged; pooling all subjects' trials into a single
ratio would weight the result by absolute EEG amplitude, which Stage~1
rescales, and would confound that rescaling with a change in physiology.

The classification windows contain no pre-cue samples, so a separate loader
takes six seconds from trial onset. BCI-IV-2a presents a fixation cross from
0 to 2\,s, the cue at 2\,s, and imagery until 6\,s; the reference window is
0.5--1.5\,s and the task window 3--6\,s. Channel indices were read from the
recorded montage. Values are Session~2 at $(0.9,0.35,0.6,0.0)$. Confidence
intervals are percentile bootstrap over the nine per-subject differences,
10{,}000 resamples, the same procedure used in Appendix~\ref{app:stats}.

\paragraph{Filter edges and smoothing.} Zero-phase filtering of a six-second
epoch rings at the edges, and the
reference window sits close enough to the start that the ringing matters. We
therefore recomputed the whole analysis with one second of reflection padding
before filtering, trimmed afterwards, and a 250\,ms moving average in time,
which is the smoothing the ERD literature applies. Every value moves by
between 1 and 3 percentage points and none changes sign or ordering: right-hand
imagery reads $-11.08\%$ at C3 and $-7.45\%$ at C4 undefended against the
$-12.02\%$ and $-8.34\%$ in Table~\ref{tab:erds}, and left-hand imagery
remains symmetric across the two sites. The table reports the unpadded values
for consistency with the operating-point record; the correlations below are
from the padded and smoothed recomputation.

The three right-hand intervals exclude zero and the three left-hand intervals
do not, so the attenuation is confined to the class that lateralises. Left-hand
imagery does not lateralise in this dataset: C4 and C3 read $-11.27\%$ and
$-11.02\%$ undefended, a difference far smaller than the between-subject
spread, so no contrast is available there to preserve or to lose.

\paragraph{Topography.} Beyond the three central channels, we computed ERD
across the full 22-channel montage per subject and per class, with the
padding and smoothing described above. The correlation between a subject's
raw and defended topography is 0.845 for right-hand imagery and 0.780 for
left, averaged over the nine subjects. On the right-minus-left contrast,
which cancels the non-specific desynchronisation both classes share, the
correlation is 0.79 [0.63, 0.92]. The raw contrast reads $-0.80\%$ at C3
against $+4.25\%$ at C4, a gradient of $-5.04$ points, and after the transform
$-1.16$ points. We report these as numbers rather than as maps. The contralateral contrast in
this dataset is small relative to the between-subject spread, one subject
having a raw value at C5 an order of magnitude above the rest, so topographic
plots of it are not informative at this effect size. The compression of the
gradient, not the preservation of its shape, is the physiologically
meaningful quantity, and it is consistent with the \bciCost-point decoding
cost.

\paragraph{Time and frequency.} The band-power analysis above fixes a
13--30\,Hz band and a 3--6\,s window. Computing ERD across 15 frequency bins
from 6 to 34\,Hz over the whole epoch, with the same padding and smoothing,
places the desynchronisation in the $\beta$ band after the 2\,s cue in both
conditions, so the transform does not shift the response in time or
frequency. Its depth is reduced: averaged over the $\beta$ band and the task
window, right-hand imagery at C3 reads $-11.08\%$ undefended and $-4.05\%$
defended, and at C4 $-7.45\%$ and $-4.94\%$. Left-hand imagery is unchanged
within a point at both sites, consistent with its lack of lateralisation in
this dataset.

\section{Full Differential Privacy Results}
\label{app:dp}
\paragraph{Formulation.} \emph{Adjacency}: two releases differ in the
covariance of a single trial. \emph{Released object}: the trial covariance,
trace-normalised to unit trace, so $\|C\|_F \le \operatorname{tr}(C) = 1$ and
$\|C\|_1 \le c$; under replacement this gives $\Delta_2 \le 2$ and
$\Delta_1 \le 2c$ with $c$ the channel count. \emph{Calibration}: Laplace
noise of scale $\Delta_1/\varepsilon$ for $\varepsilon$-DP, and Gaussian
noise of standard deviation $\Delta_2\sqrt{2\ln(1.25/\delta)}/\varepsilon$
for $(\varepsilon,\delta)$-DP at $\delta = 10^{-5}$
\citep{dwork2006calibrating,dwork2014algorithmic}. The noise matrix is drawn
symmetrically, as when releasing a covariance \citep{dwork2014analyzegauss},
and eigenvalues are clipped to keep the result positive semi-definite, which
is post-processing and consumes no budget. The adjacency relation and the
sensitivity bound are ours: neither work releases a per-trial covariance, and
the local mode is a local-DP model rather than the aggregate release those
papers analyse.

\begin{figure*}[tb]
\centering
\includegraphics[width=\textwidth]{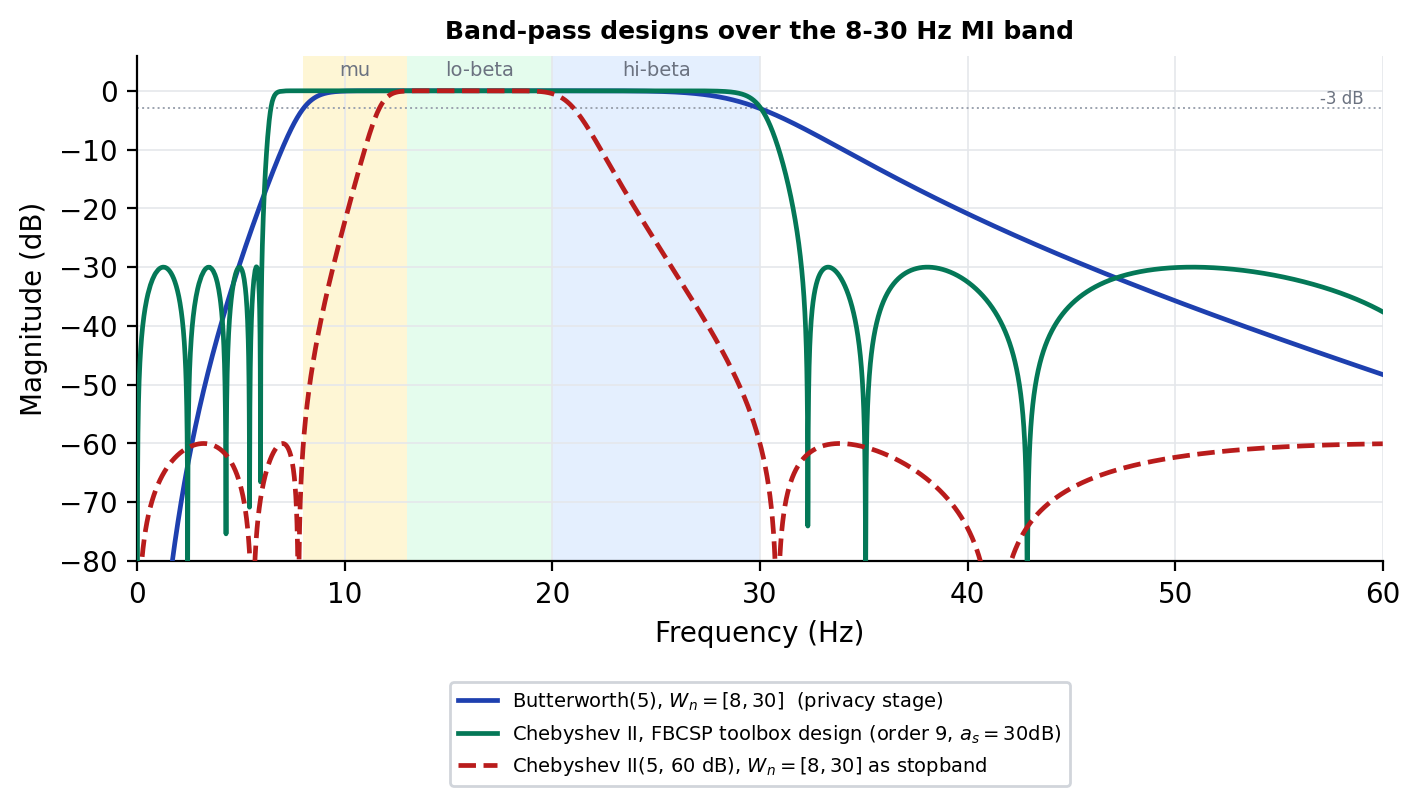}
\caption{Magnitude response over the 8--30\,Hz MI band at 250\,Hz sampling. Blue: the fifth-order Butterworth used by the privacy stages. Green: the Chebyshev~II design as specified by the reference FBCSP toolbox ($a_p{=}3$\,dB, $a_s{=}30$\,dB, 2\,Hz transition allowance, order~9 from \texttt{cheb2ord}). Red dashed: Chebyshev~II(5, 60\,dB) with the nominal band limits supplied as the stopband. The first two are equivalent across the MI band; only the third removes it.}
\label{fig:freqresponse}
\end{figure*}

Both modes suppress re-identification, and both destroy the task. Every DP
condition leaves the decoders between 29\% and 36\% against 25\% chance,
where the proposed transform leaves 64.9\%. Raising $\varepsilon$ from 1 to
10 changes nothing measurable on either axis, because at the sensitivity
these bounds require the noise dominates the signal at every setting in the
sweep: at $\varepsilon = 1$ the Laplace scale is $\Delta_1 = 44$ against
covariance entries of order $1/22$.

FBCSP could not be fitted under central mode on any subject, and the reason
is informative. At these noise levels roughly half the eigenvalues of the
noised covariance are negative, and the projection back onto the positive
semi-definite cone collapses them onto a floor, leaving a release of
effective rank 11 of 22. The generalised eigenproblem CSP solves has no
solution on such input. The three remaining decoders do not diagonalise a
class-covariance pair and are unaffected.

\begin{table}[tb]
\centering
\caption{Magnitude response (dB) at representative frequencies.}
\label{tab:filtermag}
\small
\begin{tabular}{lccc}
\toprule
Freq (Hz) & Butterworth(5) & Chebyshev~II, toolbox spec & Chebyshev~II(5, 60\,dB), band as stopband \\
\midrule
\phantom{0}8 (edge) & $-3.01$ & $\phantom{-}0.00$ & $-60.00$ \\
10 & $-0.05$ & $\phantom{-}0.00$ & $-22.20$ \\
14 & $-0.00$ & $\phantom{-}0.00$ & $-0.00$ \\
20 & $-0.00$ & $-0.00$ & $-0.49$ \\
30 (edge) & $-3.01$ & $-2.77$ & $-60.00$ \\
\bottomrule
\end{tabular}
\end{table}

\section{Filter Design: Butterworth vs.\ Chebyshev II}
\label{app:filter}
The toolbox Chebyshev~II design and the Butterworth design are
interchangeable over the band that matters here: both are within 3\,dB at the
edges and flat in between. They differ only in roll-off outside the band,
where Chebyshev~II is sharper at the cost of stopband ripple.

The 60\,dB figure that motivates the apparent conflict comes from supplying
$[8,30]$\,Hz as the \emph{stopband} rather than the passband, which
attenuates both edges by 60\,dB and removes 80--87\% of $\mu$-band power.
That is a specification error rather than a property of the filter family,
and Figure~\ref{fig:freqresponse} shows the three curves together so the
distinction is visible. We record which design each component uses so that
the spectral content seen by attacker and decoder is unambiguous.

\end{document}